\documentclass{jfm}
\usepackage{graphicx}
\usepackage{amsmath, amssymb}

\usepackage{pdflscape}
\usepackage{fancyhdr}
\fancypagestyle{lscape}{%
\fancyhf{} 
\fancyfoot[LE]{}
\fancyfoot[LO] {}
 
}
\usepackage{afterpage} 

\pdfmapfile{+rsfso.map}
\DeclareSymbolFont{rsfso}{U}{rsfso}{m}{n}
\DeclareSymbolFontAlphabet{\mathscr}{rsfso}

\usepackage{bm}
\usepackage{microtype}

\usepackage[pdfauthor={Philippe H. Trinh}]{hyperref}
\usepackage[usenames, dvipsnames]{xcolor}
\hypersetup{colorlinks=true,linkcolor=MidnightBlue,citecolor=MidnightBlue}

\usepackage{natbib}
\usepackage{array}
\usepackage{booktabs}
\usepackage{multirow}

\usepackage{tikz}

\newcommand*{\ep}{\epsilon}
\newcommand*{\F}{\mathrm{F}}
\renewcommand*{\i}{\mathrm{i}}
\newcommand*{\im}{\mathrm{i}}
\newcommand*{\e}{\mathrm{e}}
\newcommand*{\Oh}{\mathcal{O}}
\newcommand*{\bq}{\bar{q}}
\newcommand*{\bqexp}{\bar{q}_\text{exp}}
\newcommand*{\bt}{\bar{\theta}}
\newcommand*{\bG}{\bar{G}}
\newcommand*{\bPG}{\overline{\mathcal{P}G}}
\newcommand*{\bQ}{\bar{Q}}

\renewcommand*{\H}{\mathscr{H}}
\newcommand*{\bH}{\hat{\H}}
\renewcommand*{\Re}{\operatorname{Re}}
\renewcommand*{\Im}{\operatorname{Im}}

\renewcommand*{\j}{\mathrm{j}}
\renewcommand*{\k}{\mathrm{k}}
\newcommand*{\Eb}{\mathcal{E}_\mathrm{bern}}
\newcommand*{\Ebb}{\hat{\mathcal{E}}_\mathrm{bern}}
\newcommand*{\Ei}{\mathcal{E}_\mathrm{int}}

\newcommand*{\Q}{\mathcal{Q}}
\renewcommand*{\P}{\mathcal{P}}
\newcommand*{\N}{\mathcal{N}}

\newcommand*{\Iexp}{I_\mathrm{exp}}
\newcommand*{\qexp}{q_\mathrm{exp}}
\newcommand*{\texp}{\theta_\mathrm{exp}}

\newcommand*{\de}{\operatorname{d\!}{}} 
\newcommand{\dd}[2]{\frac{\de#1}{\de#2}}

\usepackage{mathabx}
\usepackage{esint} 
\def\Xint#1{\mathchoice
   {\XXint\displaystyle\textstyle{#1}}%
   {\XXint\textstyle\scriptstyle{#1}}%
   {\XXint\scriptstyle\scriptscriptstyle{#1}}%
   {\XXint\scriptscriptstyle\scriptscriptstyle{#1}}%
   \!\int}
\def\XXint#1#2#3{{\setbox0=\hbox{$#1{#2#3}{\int}$}
     \vcenter{\hbox{$#2#3$}}\kern-.5\wd0}}

\def\YYint#1#2#3{{\setbox0=\hbox{$#1{#2#3}{\int}$}
     \vcenter{\hbox{\scalebox{1}[-1]{$#2#3$}}}\kern-.5\wd0}}

\def\dashint{\Xint-}
\def\cint{\Xint\righttoleftarrow}

\title[On reduced models for gravity waves generated by moving bodies]{On reduced models for gravity waves \\ generated by moving bodies}

\author[P.H. Trinh]{P\ls H\ls I\ls L\ls I\ls P\ls P\ls E\ns H.\ns T\ls R\ls I\ls N\ls H}

\affiliation{%
Oxford Centre for Industrial and Applied Mathematics, \\ Mathematical Institute, University of Oxford, Oxford OX2 6GG, UK}

\pubyear{XXXX}
\volume{YYY}
\pagerange{xxx--xxx}
\date{\today \ [Draft]}

\newtheorem{result}{Principal Result}

\begin{document}

\maketitle

\begin{abstract}
In 1982, Marshall P. Tulin published a report proposing a framework for reducing the equations for gravity waves generated by moving bodies into a single nonlinear differential equation solvable in closed form [\emph{Proc. 14th Symp. on Naval Hydrodynamics}, 1982, pp.19--51]. Several new and puzzling issues were highlighted by Tulin, notably the existence of weak and strong wave-making regimes, and the paradoxical fact that the theory seemed to be applicable to flows at low speeds, \emph{``but not too low speeds"}. These important issues were left unanswered, and despite the novelty of the ideas, Tulin's report fell into relative obscurity. Now thirty years later, we will revive Tulin's observations, and explain how an asymptotically consistent framework allows us to address these concerns. Most notably, we will explain, using the asymptotic method of steepest descents, how the production of free-surface waves can be related to the arrangement of integration contours connected to the shape of the moving body. This approach provides an intuitive and visual procedure for studying nonlinear wave-body interactions.
\end{abstract}

\section{Introduction} \label{sec:intro}

The motivation for this paper stems from an important, but seemingly forgotten 1982 report by Prof. Marshall P. Tulin presented during the $14^\text{th}$ Symposium on Naval Hydrodynamics, titled ``\emph{An exact theory of gravity wave generation by moving bodies, its approximation and its implications}" \citep{tulin_1982_an_exact}. Some thirty years after its publication, Tulin wrote of his original motivation for pursuing the former study:
\begin{quotation}
\noindent \emph{What were the relations and connections among these various nonlinear approximations---ray, ``slow ship,", second order, formal straining, and Guilloton---that had arisen by the end of the 1970s? [...] I had earlier in the 1970s become intrigued by the Davies transformation of the nonlinear free-surface problem, which was revealed in Milne-Thompson's legendary banquet speech [in 1956]. My hope was that my extension of the Davies theory would provide an exact result in analytical form, which even in its complexity could then be subject to various approximations, the connections of which could thereby be discerned. And so it turned out.} \citep[p.242]{tulin_2005_reminiscences_and}
\end{quotation}

In the 1982 paper, Tulin sought to derive a rigorous mathematical reduction of the water wave equations in such a way that certain nonlinear contributions within the free surface equations could be preserved. The resultant model was analytically simple, and took the form of a single complex-valued linear differential equation. The theory was also powerful, and provided a formulation that could relate the geometry of a moving body directly with the resultant free-surface waves.

However, several important and surprising issues were raised by Tulin regarding the model and its interpretation, and in particular, he had noted a paradoxical behaviour of the model at low speeds. In the years that followed, perhaps owing to the difficulty of the model's derivation, Tulin's fundamental questions were never re-addressed. In this paper, we shall present an asymptotically consistent derivation that corrects Tulin's model, and puts to rest many of the issues previously highlighted. 

More specifically, we shall present an explicit solution written in terms of a single integral that properly describes the form of water waves produced by two-dimensional moving bodies at low speeds. Then, by applying the asymptotic method of steepest descents, we are able to observe how the production of free-surface waves will change depending on the deformation of integration contours connected to the geometry of the moving body. This approach provides an intuitive and visual procedure for studying wave-body interactions.

\subsection{Tulin's water wave model} \label{sec:subtulin}

The essential derivation behind Tulin's model begins from Bernoulli's equation applied to a free surface with streamline, $\psi = 0$, 
\begin{equation} \label{bern}
\frac{\ep}{q} \dd{q}{\phi} + \frac{\sin \theta}{q^3} = 0, 
\end{equation}
where $q$ is the fluid speed, $\theta$ the streamline angle, $\phi$ the potential, and the non-dimensional parameter
\begin{equation}
\ep = \F^2 = \frac{U^2}{gL}
\end{equation}
is the square of the Froude number for upstream speed $U$, gravity $g$, and length scale $L$. If the sinusoidal term is split according to the identity
\begin{equation} \label{sinreduce}
	3\sin \theta = \sin 3\theta + 4\sin^3 \theta,
\end{equation}
then \eqref{bern} can be written in complex-valued form
\begin{equation} \label{bern2}
\Re\left\{\frac{\ep}{G} \dd{G}{w} - \frac{\im}{G}  + \P(w, q, \theta)\right\}  = 0,
\end{equation}
where $G = (q\e^{-\im\theta})^3$ is an analytic function of the complex potential, $w = \phi + \im \psi$, and the above is evaluated on $\psi = 0$ where $\Re(\P) = 4\sin^3 \theta/q^3$. The rather curious substitution of \eqref{sinreduce} is attributed to \cite{davies_1951_the_theory}, who had argued that if $\P$ is considered to be small, then \eqref{bern2} yields a linearized version of Bernoulli's equation (in $G$) that preserves the essential nonlinearity (in $q^3$) describing the structure of steep gravity waves. 

Inspired by this idea, Tulin considered the extension to general free surface flows over a moving body. Since the function in the curly braces of \eqref{bern2} is an analytic function in $w$, except at isolated points in the complex plane, then analytic continuation implies 
\begin{equation} \label{bern3}
\left\{ \frac{\ep}{G} \dd{G}{w} - \frac{\im}{G}  + \P(w, q, \theta)\right\}  = -\Q(w),
\end{equation}
where by matching to uniform flow, $G \to 1$ and thus $\Q \to \i$ as $w \to -\infty$. The function $\Q(w)$ is purely imaginary on $\psi = 0$. In the case of flow without a body or bottom boundary, $\Q \equiv \i$, but otherwise, $\Q$ will encode the effect of the obstructions through its singular behaviour in the complex plane. If the nonlinear contribution, $\P$, is neglected in \eqref{bern3}, then the exact solution can be written as 
\begin{equation}
  G(w) = \left[\frac{\i }{\ep} \int^w \e^{\frac{1}{\ep} \int^t \Q(s) \, \de{s}} \, \de{t} \right]\e^{-\frac{1}{\ep} \int^w \Q(s) \, \de{s}}.
\end{equation}

This process seems to yield the exponentially small surface waves at low speeds. However, Tulin noted that as $\ep \to 0$, there would be locations on the free surface where $\Q = 0$, and this would lead to unbounded steepness. He wrote of
\begin{quotation}
\noindent \emph{\ldots the revelation that for sufficiently strong disturbances \ldots waves arise at discrete points on the free surface which\ldots do not become exponentially small with decreasing Froude number, but rather tend to unbounded steepness as $\F \to 0$}.
\end{quotation}
He thus proposed the following result:
\begin{quotation} \label{tulinquote}
\noindent \emph{The most important comment to make is that for given $\ep$, no matter how small, this so-called low speed theory is not valid for sufficiently low speeds. It is a theory valid for low, but not too low speeds!}
\end{quotation}
The matter was left at that, and in the three decades following Tulin's ingenious paper, the peculiarities surrounding the asymptotic breakdown of \eqref{bern3} were never directly re-addressed (though we mention the paper by \cite{vanden-broeck_1995} which develops numerical solutions for the case $\Q = \text{constant}$ and $\P \neq 0$). 

Note that in addition to the investigation in the limit $\ep \to 0$, Tulin had also intended to produce an \emph{exact} reduction, where the $\P$ term in \eqref{bern3} was handled through a theoretically posited nonlinear coordinate transformation. However, it is never clear how this transformation is used or derived in practice, except through nonlinear computation of the full solutions. 

\subsection{E.O. Tuck's linearized integral equation} \label{sec:introtuck}

Independently from Tulin, E.O. Tuck later presented a series of papers \citep{tuck_1990_water_non-waves, tuck_1991_ship-hydrodynamic_free-surface, tuck_1991_waveless_solutions} where he attempted to distill the wave-making properties of wave-body problems into a linear singular equation. The equation presented [eqn (22) of \cite{tuck_1991_ship-hydrodynamic_free-surface}] was
\begin{equation} \label{tucksimp}
\epsilon \H \dd{y}{\phi} + y = \mathcal{F}(\phi),
\end{equation}
where $y(\phi)$ is the height of the free surface, $\mathcal{F}(\phi)$ is a function related to the moving body, and $\H$ is a integral operator known as the Hilbert transform (to be introduced in \S\ref{sec:form}). The difficulty in solving \eqref{tucksimp} is that the Hilbert transform is a \emph{global} operator, requiring values of $y$ over the entire domain.


In the \citeyear{tuck_1991_ship-hydrodynamic_free-surface} work, Tuck explained how the action of $\H$ could be viewed as similar to that of the differential operator $\dd{}{\phi}$. This reduction was mostly pedagogic in nature, but Tuck was motivated by the fact that $\H$ and the differential operator will act similarly in the case of sinusoidal functions. He went on to study the various properties of the singular differential equation 
\begin{equation} \label{tucktoy}
\ep \dd{^2 y}{\phi^2} + y = \mathcal{F}(\phi),
\end{equation}
that depend on the specification of the `body' given by $\mathcal{F}(\phi)$. 

Apparently, Tuck had been unaware of Tulin's (\citeyear{tulin_1982_an_exact}) work, but a chance meeting of the two occured during a conference leading to the publication of \cite{tuck_1991_waveless_solutions}. We are fortunate enough to possess the archived questions of the meeting, where we discover that Tulin had asked the following question:
\begin{quotation}
\noindent \emph{Isn't it true that the two dimensional wavemaker problem can be presented in terms of an ordinary differential equation in the complex domain, at least to some higher order of approximation?}
\end{quotation}
Tuck replied that he was unsure of the generality of the reduction to problems including different geometries, but noted the connection to the \cite{davies_1951_the_theory} reduction \eqref{sinreduce}:
\begin{quotation}
\noindent \emph{I do not know the answer to this question...My ``F(x)" in some way represents a very general family of ``wavemakers", with structure in both spatial dimensions, and I have doubts as to whether the problem can then be converted (exactly) to a differential equation. On the other hand, a few years ago I in fact used the method that you describe, and it is associated with the approximation $\sin\theta \approx \frac{1}{3}\sin3\theta$. }
\end{quotation}
Although Tuck's toy reduction \eqref{tucktoy} should only be regarded as illustrative (the governing differential equation should rather be first order), what is apparent in his work is the desire to systematically reduce both the nonlinearity of Bernoulli's equation, and the global effect of $\H$ into a single ordinary differential equation. In particular, it is Tuck's search for a reduction of the operator, $\H$, that was missing from earlier works on this topic (including \citealt{tulin_1982_an_exact}).

\subsection{Other works and reductions} \label{sec:otherworks}

Certainly, Tulin and Tuck were not the only ones to seek simpler reductions or formulations of the nonlinear equations for wave-body interactions, and indeed, Tulin relates his work to the integral models proposed by \cite{inui_1977_a_study} and \cite{dawson_1977_a_practical}. Reviews of these and other models were presented by \cite{doctors_1980_comparison_of} and \cite{miloh1985study}, and many others. However, what distinguishes Tulin and Tuck's work is the shifted focus towards the analytic continuation of the flow problem into the complex domain; indeed, Tulin notes his strong motivation by the work of \cite{davies_1951_the_theory} in his \citeyear{tulin_2005_reminiscences_and} review.

As we have noted in \S\ref{sec:subtulin}, the low-Froude or low speed limit of $\ep \to 0$ is the essential approximation in which analytical results can be derived. The subtleties of studying this singular limit can be traced back to a seminal report by \cite{ogilvie_1968_wave_resistance:}, who detailed certain oddities with the previously developed analytical approximations of free surface flow past a submerged body. Chief amongst such oddities was the fact that the individual terms of a series approximation in powers of $\ep$ would fail to predict surface waves. Thus, one might decompose the surface speed into a regular series expansion and an error term, $\bq$, with
\begin{equation} \label{qseriestest}
   q(\phi) \sim \Bigl[q_0(\phi) + \ep q_1(\phi) + \ldots + \ep^{N-1} q_{N-1}(\phi)\Bigr] + \bq.
\end{equation}
The challenge, Ogilvie realized, was that water waves were exponentially small in $\ep$, with $\bq = \Oh(\e^{-\text{const.}/\ep})$, and thus \emph{beyond-all-orders} of any individual term of the regular series. 

By linearizing about the zeroth order solution, $q = q_0 + \bq$ for $\bq \ll 1$, and strategically perserving certain terms in Bernoulli's equation, Ogilvie developed a general analytical approximation for the exponentially small surface waves. The approximation, however, was not asymptotically consistent, and the search for a complete numerical and analytical treatment of the low Froude limit would inspire many papers in the subsequent years. 

One of the key issues we will explore in this paper addresses the question of how many terms must be included in the linearization of \eqref{qseriestest} in order to obtain the exponential. Originally \cite{ogilvie_1968_wave_resistance:} had chosen $N = 1$, but later revised to $N = 2$ in \cite{ogilvie_1982_water_waves} (who quoted the work of \cite{dagan_1972_two-dimensional_free-surface} and in particular, the study by \cite{keller_1979a} in applying the WKB method to streamline ship waves). 


\begin{table} 
\begin{center}\footnotesize
\begin{tabular}{>{\raggedleft\arraybackslash}p{5.3cm}cp{7cm}}
\emph{Historical significance} && \emph{Papers} \\[0.5\baselineskip]
Origin of the low-Froude paradox && \cite{ogilvie_1968_wave_resistance:} \\[0.5\baselineskip]
Two-dimensional and three-dimensional linearizations &&
\cite{dagan_1972_two-dimensional_free-surface},
\cite{keller_1979a},
 \cite{ogilvie_1982_water_waves},
 \cite{doctors_1980_comparison_of},
\cite{tulin_1984_surface_waves},
 \cite{brandsma_1985a}
 \\[0.5\baselineskip]
On numerical solutions && 
\cite{vanden-broeck_1977_computation_of},
\cite{vanden-broeck_1978_divergent_low-froude-number},
\cite{madurasinghe_1986_ship_bows},
\cite{farrow_1995_further_studies}
\\[0.5\baselineskip]
On exponential asymptotics applied to water waves && 
 \cite{chapman_2002_exponential_asymptotics, chapman_2006_exponential_asymptotics}, 
 \cite{trinh_2011_do_waveless}, 
 \cite{trinh_2013_new_gravity-capillary, trinh_2013a_new_gravity-capillary}, 
 \cite{lustri_2013_exponential_asymptotics,lustri_2014a}, 
 \cite{trinh_2014_the_wake} \\[0.5\baselineskip]
Review articles && \cite{tuck_1991_ship-hydrodynamic_free-surface}, \cite{tulin_2005_reminiscences_and} \\[0.5\baselineskip]
\end{tabular}
\end{center}
\caption{The development of the low-Froude paradox. \label{tab:res} }
\end{table}

We shall not pursue, in great detail, the history of the low Froude problem that followed Ogilvie's seminal report, but instead refer to the review papers by \cite{tuck_1991_ship-hydrodynamic_free-surface} and, particularly, \cite{tulin_2005_reminiscences_and} where certain aspects of the low Froude difficulty are discussed. Additional historical details are presented in \S1 of \cite{trinh_2011_do_waveless}, and a selection of papers on the problem is presented in Table~\ref{tab:res}. The method we apply in this paper, which combines an approach of series truncation with the method of steepest descents, is unique from the previously listed works.

\subsection{Goals of this paper} \label{sec:goals}

We have three principal goals in this work. 

\emph{(i) We wish to demonstrate how Tulin's formulation \eqref{bern3} can be systematically derived and studied in the low speed limit.} The source of Tulin's puzzling results can be resolved through better understanding of two effects: first, the \emph{ad-hoc} linearization of the nonlinear $\P$ function; and second, the role of the forcing function, $\Q(w)$. We clarify these details and demonstrate, through numerical solutions of the full nonlinear water wave equations, the convergence of different proposed models in the limit $\ep \to 0$. 

Let $\bq$ be the fluid speed corresponding to the water waves, as in \eqref{qseriestest}. One of our principal results (presented on p.~\pageref{result:integro}) is to demonstrate that the exponentially small waves, $\qexp$, are described to leading order by the first-order equation,
\begin{equation} \label{qbareqtest}
  \ep \bq' + \biggl[ \frac{\im\j}{q_0^3} + \ep \left(\frac{2 q_0'}{q_0} - \frac{3\im \j q_1}{q_0^4}\right) \biggr]\bq 
= R(w; \bH[\bt]),
\end{equation}
where $\j = \pm 1$ and $\qexp$ is then given by the sum of $\bq$ and its complex conjugate. Different choices of the series truncation $N$ yield different versions of the right-hand side of \eqref{qbareqtest}, but only change the predicted wave amplitudes by a numerical factor. The leading order $q_0$ contains the prescription of the moving body, and can thus be related to Tuck's $\mathcal{F}$ function in \eqref{tucksimp}. The formulation in terms of the speed, $q$, rather than Tulin's combined function $G = (qe^{-\im\theta})^3$ in \eqref{bern3} is more natural, but we will relate Tulin's equation to our own in \S\ref{sec:tulinconnect} and Appendix~\ref{sec:tulinconnect2}.

\emph{(ii) We also study the associated integral form of the solution using the method of steepest descents.} We shall demonstrate how the appearance of surface waves can be associated with sudden deformations of the integration contours if the solution to \eqref{bern3} is analytically continued across critical curves (Stokes lines) in the complex plane. This process is known as the Stokes Phenomenon \citep{berry_1991_asymptotics_superasymptotics,olde-daalhuis_1995_stokes_phenomenon,trinh_2010_asymptotic_methods_incol}. The novelty of this steepest descents methodology is that it will allow us to not only relate the surface waves directly to the geometry of the moving body, but it will also allow us to observe how integration contours change depending on the geometry of the obstruction. In particular, we conclude that provided there exists a solution of the full potential flow problem, there are no issues with the approximations as $\ep \to 0$ limit.  

\emph{(iii) Our last goal is to provide a link between the Tulin approximation \eqref{bern3}, our corrected model, and also the current research on low-Froude approximations.} 

Let us now turn to a brief summary of the history of the low Froude problem.

\section{Mathematical formulation} \label{sec:form}

\begin{figure} \centering
\includegraphics[width=1.0\textwidth]{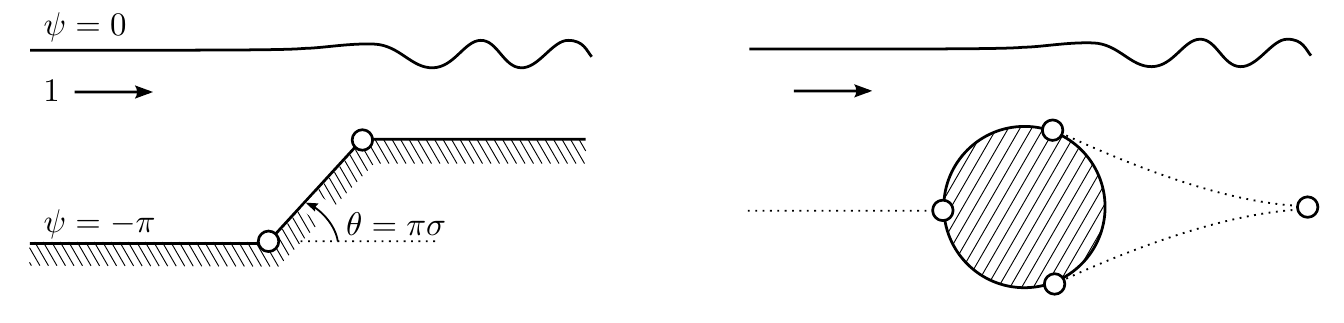}
\caption{Non-dimensional flow over an angled step (left) and flow past a circular cylinder (right). The dimensionalization is discussed in \S\ref{sec:bdint}. Flows past closed bluff objects can be complicated through wake separation (seen right), for which the nature of the separation points is unclear. \label{formstepcyl}}
\end{figure}

Let us consider steady irrotational two-dimensional flow past a moving body in the presence of gravity, $g$. The body is associated with a length scale $L$, and moves at constant velocity $U$. For instance, this body may correspond to an obstacle at the bottom of the channel (Fig.~\ref{formstepcyl}, left), a submerged object (Fig.~\ref{formstepcyl}, right), or a surface-piercing ship (Fig.~\ref{formship}, left). We shall state more precise geometrical restrictions in \S\ref{sec:bdint}.

The velocity potential, $\phi$, satisfies Laplace's equation in the fluid region, $\nabla^2 \phi = 0$. On all boundaries, the kinematic condition implies the normal derivative is zero, $\partial \phi/\partial n = 0$, while on the free surface, Bernoulli's equation requires that
\begin{equation}
	\frac{1}{2} \left( |\nabla \phi|^2 - U^2\right) + gy = \text{const.}
\end{equation}
Subsequently, all quantities are non-dimensionalized using the velocity and lengths scales, $U$ and $L$, and we introduce a Cartesian coordinate system $(x,y)$ such that the body is fixed in the moving frame of reference.

With $z = x + \im y$, we define the complex potential, $w = \phi + \im \psi$, and the complex velocity is given by, 
\begin{equation} \label{compvel}
\dd{w}{z} = q\e^{-\im \theta}.
\end{equation}
Here $\psi$ is the stream function, $q$ is the fluid speed, and $\theta$ is the streamline angle, measured from the positive $x$-axis. Without loss of generality, we choose $\psi = 0$ on the free surface and $\psi < 0$ within the fluid. We define the logarithmic hodograph variable,
\begin{equation} \label{hodograph}
  \Omega = \log q - \im \theta,
\end{equation}
and seek to solve the flow problem inversely, that is, use $w$ as an independent variable and seek $\Omega = \Omega(w)$ in the lower-half $w$-plane. The advantage of this hodograph formulation is that with the free surface at $\psi = 0$, its position is known in the $(\phi, \psi)$-plane, even though its shape in the physical $(x,y)$-plane is unknown. Differentiating Bernoulli's equation with respect to $\phi$ and using \eqref{compvel} then yields the non-dimensionalized form \eqref{bern}. 


\subsection{Boundary integral formulation and geometrical examples} \label{sec:bdint}


\begin{figure} \centering
\includegraphics[width=1.0\textwidth, clip=true, trim=0 0 0 3.5cm]{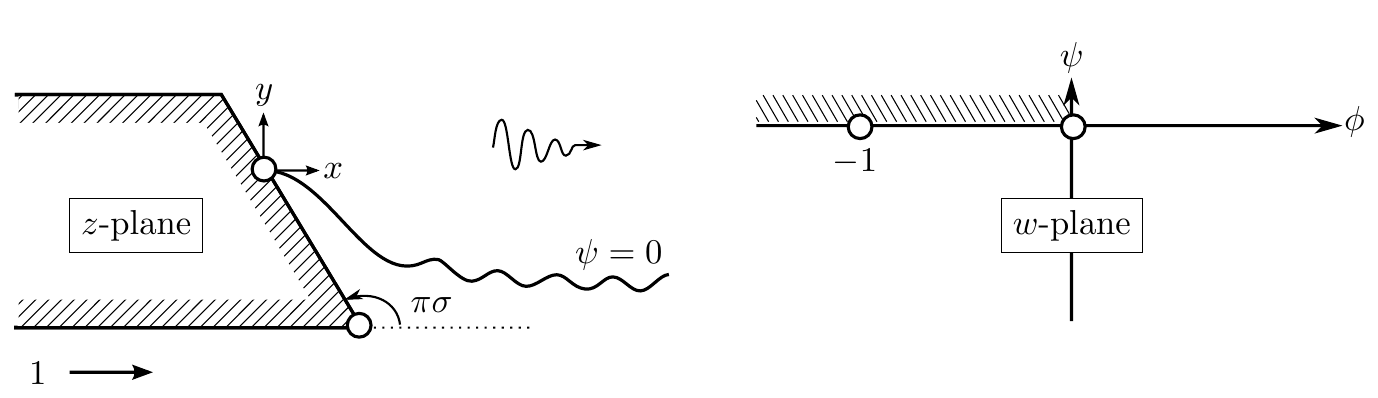}
\caption{Physical (left) and potential $w$-plane (right) for the one-cornered ship. \label{formship}}
\end{figure}

In theory, the methods we present in this paper will apply to most general two-dimensional free-surface flows. In practice, however, in order to make analytical progress, we will be constrained by problems in which the geometry of the body is known through the specification of the angle, $\theta$, in terms of the complex potential, $w = \phi + \im \psi$. 

As specific examples, let us focus on two representative geometries: (a) flow past a localized obstruction on a channel bottom (Fig.~\ref{formstepcyl}, left), and (b) flow past a semi-infinite surface piercing ship (Fig.~\ref{formship}, left). 


For flow past a varying channel bottom with dimensional upstream depth $h$, we select the length scale $L = h/\pi$, so that the flow region in the $w$-plane consists of an infinite strip between $0 \leq \psi \leq \pi$. The strip is then mapped to the upper-half $\zeta$-plane, using 
\begin{equation}\label{zetamap}
  \zeta = \xi + \i \eta = \e^{-w},
\end{equation}
where under \eqref{zetamap}, the free surface is mapped to $\xi \geq 0$, the channel bottom to $\xi \leq 0$, and the flow region to $\eta \geq 0$. 

For flows past a semi-infinite surface piercing body, we can choose the length scale to be $L = K/U$, where $K$ is a representative scale of the potential along the body (see (2.3) in \cite{trinh_2014_the_wake} for further details). We assume that the free surface attaches to the body at a stagnation point, chosen without loss of generality to be $\phi = 0$. In the potential plane, the flow region consists of $\psi \leq 0$. On the boundary $\psi = 0$, $\phi > 0$ corresponds to the free surface and $\phi < 0$ to the solid body. Since the flow is already contained within a half-plane, we do not need a further $w \mapsto \zeta$ transformation, but we shall set $\zeta = w$ so as to use the same notation. 

Bernoulli's equation \eqref{bern} provides an explicit relationship between the real and imaginary parts of $\Omega(w)$ on the free surface. However, because $\Omega(w)$ is analytic in the upper or lower-half $\zeta$-plane and tends to zero far upstream, there is a further Hilbert-transform relationship between its real and imaginary parts. Applying Cauchy's theorem to $\Omega(w)$ over a large semi-circle in the upper (channel flow) or lower (surface-piercing flow) half-planes, and taking the real part gives the Principal Value Integral,
\begin{equation} \label{firstbdint}
  \log q = \frac{\j}{\pi} \dashint_{-\infty}^{\infty} \frac{\theta(\xi')}{\xi' - \xi}  \, \de{\xi'} \quad \text{for $\xi \in \mathbb{R}$,}
\end{equation}
where $\j = -1$ for flow in the channel configuration and $\j = 1$ for flow in the ship configuration. The range of integration can be split into an evaluation over the solid boundary and over the free surface. We write
\begin{equation} \label{bdintgen}
\log{q} = \log q_s + \j\H[\theta](\xi) \quad \text{for $\xi \in\mathbb{R}$},
\end{equation}
where $\H$ denotes the Hilbert transform operator on the semi-infinite interval, 
\begin{equation}
\H[\theta](\xi) = \frac{1}{\pi} \dashint_{0}^{\infty}
\frac{\theta(\xi')}{\xi' - \xi} \ \de{\xi'}.
\end{equation}

We assume that for a given physical problem, the angle $\theta$ is known for the particular geometry, and thus the function $q_s$ that appears in \eqref{bdintgen} is known through calculating
\begin{equation} \label{logqs}
\log q_s = \frac{\j}{\pi} \int_{-\infty}^0 \frac{\theta(\xi')}{\xi' - \xi} \ \de{\xi'}
\end{equation}
Note that it is somewhat misleading to describe $\theta(\xi)$ as ``known" for $\xi < 0$ since in practice, we would specify $\theta$ as a function of the physical coordinates $(x,y)$. However, specifying different forms of $\theta$ as a function of $\xi$ (or the potential, $\phi$) is typically sufficient to obtain the qualitatively desired geometry shape.  

For instance, we consider the step geometry, 
\begin{equation} \label{steptheta}
	\theta_\textrm{step} = \begin{cases}
	0 & \xi \in (-\infty, -b) \cup (-a, 0)\\ 
	\pi\sigma & \xi \in (-b, -a)
	\end{cases}
\end{equation}
where $0 < a < b$, which corresponds to a step of angle $\pi\sigma$. Such topographies have been considered by \cite{king_1987_free-surface_flow}, \cite{chapman_2006_exponential_asymptotics}, \cite{lustri_2012_free_surface}, others. In this case, \eqref{logqs} and \eqref{zetamap} yields
\begin{equation} \label{qs_step}
  q_s = \left(\frac{\xi + b}{\xi + a}\right)^\sigma = \left(\frac{\e^{-\phi} + b}{\e^{-\phi} + a}\right)^\sigma.
\end{equation}

Similarly, a semi-infinite ship with a single corner of angle $\pi \sigma$ can be specified using 
\begin{equation} \label{shiptheta}
	\theta_\textrm{ship} = \begin{cases}
	0 & \phi \in (-\infty, -1) \\ 
	\pi\sigma & \phi \in (-1, 0).
	\end{cases}
\end{equation}
Such two-dimensional hull shapes have been considered in the works of \cite{vanden-broeck_1978_divergent_low-froude-number}, \cite{farrow_1995_further_studies}, \cite{trinh_2011_do_waveless}, and others. Choosing the dimensional length, $L = K/U$ where $K$ is the value of the potential at the corner sets its non-dimensional position to $\phi = -1$. Then using \eqref{logqs}, the $q_s$ function is given by 
\begin{equation} \label{qs_ship}
  q_s = \left(\frac{\xi}{\xi + 1}\right)^\sigma = \left(\frac{\phi}{\phi + 1}\right)^\sigma,
\end{equation}
where recall the ship problem does not require an additional mapping to the half plane, so we can set $\zeta = w$ to preserve the notation.  


In this paper, we will only consider geometries, as specified through $q_s$ in \eqref{logqs}, which contain strong singularities in the complex plane---that is, poles or branch points. For instance, the step \eqref{qs_step} and ship \eqref{qs_ship} contain singularities at solid corners and stagnation points. We will see in \S\ref{sec:steep} that these singularities are often responsible for the creation of the surface waves. Weaker singularities, such as for the case of flow past a smoothed hull with a discontinuity in the curvature will be the subject of a forthcoming paper (see also \S{7} of \citealt{trinh_2014_the_wake}). 

Flows past closed objects also present challenging cases for study because typically, the geometry is not known as a function of the potential variables, and there are further difficulties with the prediction of wake separation points. For instance, potential flow over a circular cylinder (Fig.~\ref{formstepcyl}, right) was computed without a free surface by \cite{hocking2008effect} where it was shown that the properties of the separation points will vary depending on the Froude number. In theory, the methodology discussed in this paper can be applied to these challenging flows, but may require hybrid numerical-asymptotic treatments. We shall return to discuss such cases in \S\ref{sec:discuss}.

\section{The failure of the typical asymptotic expansion} \label{sec:baseasym}

In this section, we demonstrate that the regular expansion of the solution of Bernoulli's equation \eqref{bern} and the boundary integral \eqref{bdintgen} is divergent, and moreover fails to capture the free surface waves. 

In the limit $\ep \to 0$, we substitute a regular asymptotic expansion, $q = q_0 + \ep q_1 + \Oh(\ep^2)$ and $\theta = \theta_0 + \ep \theta_1 + \Oh(\ep^2)$ into the two governing equations, and obtain for the first two orders,
\begin{subequations} \label{asym01}
\begin{alignat}{3}
q_0 &= q_s, &\qquad \theta_0 &= 0, \label{asym0} \\
q_1 &= \j q_0\H[\theta_1], &\qquad \theta_1 &= -q_0^2 \dd{q_0}{\phi}. \label{asym1}
\end{alignat}
\end{subequations}

Thus, at leading order, the free surface is entirely flat, $\theta_0 = 0$, and this solution is known as the rigid wall solution. The leading-order speed, $q_0 = q_s$, is given by \eqref{qs_step} for the step, or \eqref{qs_ship} for the ship, but fails to capture the surface waves. Because all the subsequent orders in the asymptotic scheme depend on derivatives of the leading-order solution, it stands to reason that it is impossible to encounter a sinusoidal term, despite going to any algebraic order in $\ep$.

Using the numerical algorithms outlined \cite{trinh_2011_do_waveless}, we calculate the numerical solutions of \eqref{bern} and \eqref{bdintgen} for the case of a rectangular ship, with $\sigma = 1/2$ and at $\ep = 1.0$ in Fig.~\ref{fig:profilestern}. Indeed, it is seen that the leading-order asymptotic approximation fails to capture the wave-like behaviour, but provides a good fit to the mean speed. More details of these numerical solutions will be discussed in \S\ref{sec:numerics}.

\afterpage{
\clearpage
\begin{landscape}
\thispagestyle{lscape}
\pagestyle{lscape}
\begin{figure}
\includegraphics[angle=90]{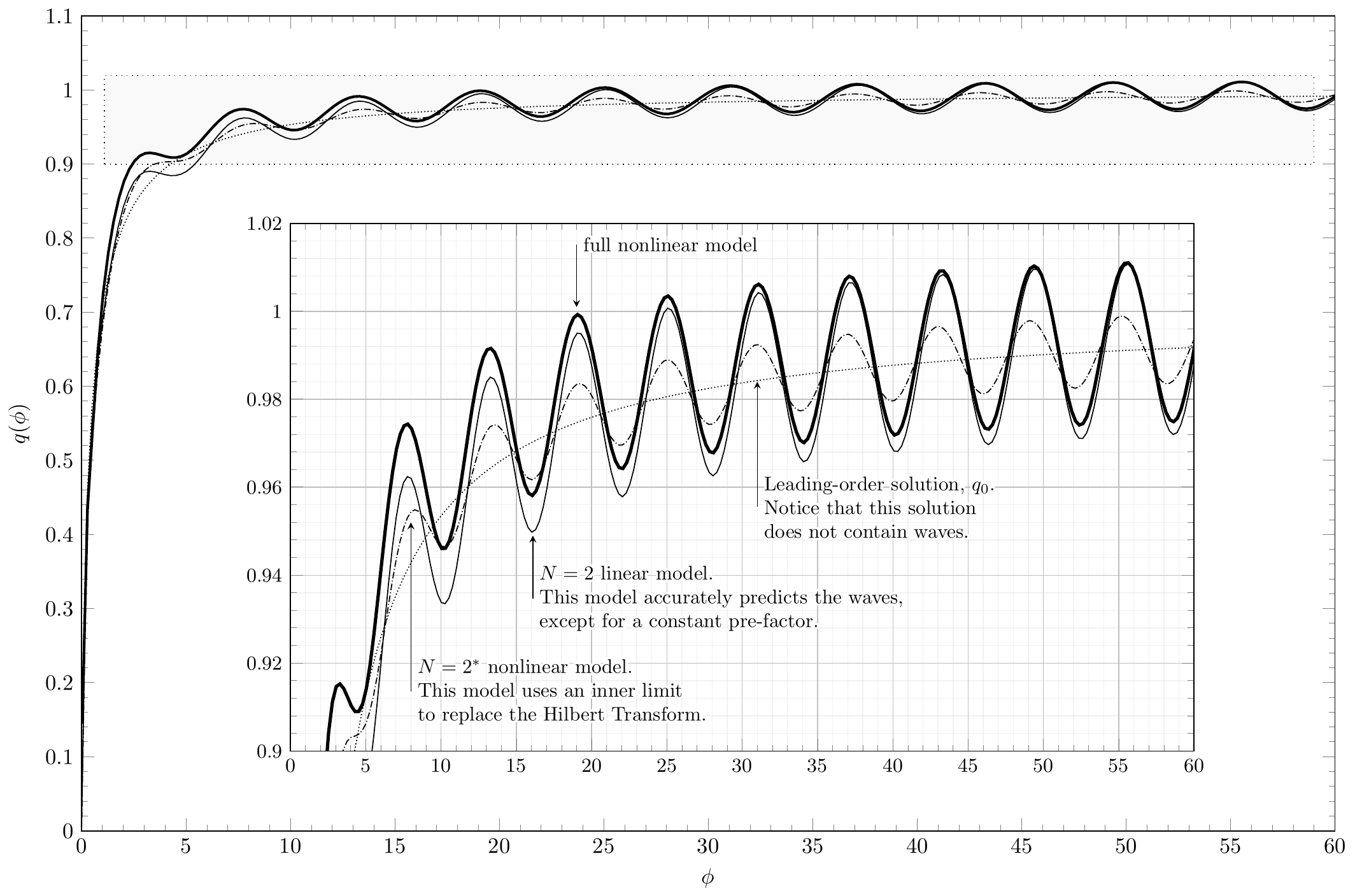}
\caption{Numerical solution (thick line) of the full stern-flow problem with $\ep = 1.0$ and $\sigma = 1/2$. The dotted line is the leading-order approximation $q_0$. Also shown are the solutions, $q_0 + 2\Re(\bq)$, from the truncated linear $N = 2$ model (thin line) and the truncated nonlinear $N = 2^*$ model (dash-dotted line). These models are discussed in \S\ref{sec:numerics} and summarized in Table~\ref{tab:models}. \label{fig:profilestern}}
\end{figure}
\end{landscape}
}

\subsection{Divergence of the asymptotic approximation}

The leading-order approximation, $q_0$, contains singularities in the analytic continuation of $\phi$ off the free surface, and into the complex plane, $\phi \mapsto \phi_r + \i \phi_c$. For instance, in the case of the step geometry \eqref{qs_step}, we see that $q_0$ contains branch points at $\phi = -a$ and $\phi = -b$, which correspond to the corner and stagnation points. 

Because of the singularly perturbed nature of Bernoulli's equation in the limit $\ep \to 0$, we can see that at each order in the asymptotic procedure, the calculation of $q_n$ and $\theta_n$ will depend on differentiation of the previous order, $q_{n-1}$ and $\theta_{n-1}$. Since the leading-order approximation is singular, this has the effect of increasing the power of the singularities with each subsequent term in the approximation. Thus in the limit that $n \to \infty$, $q_n$ and $\theta_n$ will diverge. As argued in the work of \emph{e.g.} \cite{chapman_2006_exponential_asymptotics}, the divergence at $
\Oh(\ep^n)$ is captured through a factorial over power ansatz,
\begin{equation} \label{qt_fact}
  q_n \sim \frac{Q(w) \Gamma(n + \gamma)}{\chi(w)^{n+\gamma}} \quad \text{and} \quad
  \theta_n \sim \frac{\Theta(w) \Gamma(n + \gamma)}{\chi(w)^{n+\gamma}}.
\end{equation}
where $\chi = 0$ at the particular singularity. If there are multiple singularities contributing to the divergence (see \cite{trinh_2014_the_wake} for an example), then we must include a summation over similar ansatzes of the form \eqref{qt_fact} where $\chi = 0$ at each individual singularity. There exist other cases where a more general form of the divergence is required, and this is documented in \cite{trinh_2015_exponential_asymptotics}. 

In order to derive the components, $Q$, $\chi$, and $\gamma$, we can examine the $\Oh(\ep^n)$ form of the system, and take the limit $n \to \infty$. In this paper, the late-orders behaviour is not a crucial part of the analysis, but we collect the functional forms of the components in Appendix \ref{sec:divergence}.

\section{Analytic continuation of the governing equations} \label{sec:analytic}

In \S\S\ref{sec:reduce} and \ref{sec:steep}, we will demonstrate that the leading-order water waves can be described by an integral equation, which can then be approximated by deforming the path of integration into the complex plane. Thus it will be necessary for us to study the analytic continuation of the free surface quantities, $q(\xi, 0)$ and $\theta(\xi, 0)$ into the complex $\xi$-plane (or analogously, the complex $\phi$-plane). 

Let us seek the complexified versions of Bernoulli's equation \eqref{bern} and the boundary integral \eqref{bdintgen}. We set $\xi \mapsto \xi_r + \im \xi_c$. For analytic continuation into the upper half-plane, we can verify that in the limit $\xi_c \to 0$ from the upper half-plane, 
\begin{equation} \label{hilbertsplit}
\frac{1}{\pi} \int_{0}^{\infty}
\frac{\theta(\xi')}{\xi' - (\xi_r + \im \xi_c)} \ \de{\xi'} = \frac{1}{\pi} \dashint_{0}^{\infty}
\frac{\theta(\xi')}{\xi' - \xi_r} \ \de{\xi'}
+ \frac{1}{\pi} \landdownint
\frac{\theta(\xi')}{\xi' - \xi_r} \ \de{\xi'}
\end{equation}
where the second integral on the right hand-side corresponds to the counterclockwise integral around $\xi' = \xi$ along a small semi-circle. An analogous argument applies for analytic continuation into the lower half-plane. Writing $\xi_r + \im \xi_c = \zeta$, we have that the analytic continuation of the Hilbert Transform, $\H$, is given by
\begin{equation} \label{Hrelate}
	\H[\theta](\zeta) = \bH[\theta](\zeta) - \im \k \theta(\zeta), 
\end{equation}
where $\k = 1$ for analytic continuation into the upper half-plane, $\k = -1$ for the continuation into the lower half-plane, and we have introduced the notation $\bH$ for the integral
\begin{equation}
 \bH[\theta](\zeta) = \frac{1}{\pi} \int_{0}^{\infty}
\frac{\theta(\xi')}{\xi' - \zeta} \ \de{\xi'}.
\end{equation}
Thus, the analytic continuation of the boundary integral equation \eqref{bdintgen} is given by 
\begin{equation} \label{intanal}
\log q + (\j\k) \im \theta = \log q_s + \j \bH[\theta]	
\end{equation}

There is a somewhat subtle aspect of analytically continuing $q$ and $\theta$ off the free surface, and the relationship of these quantities with the physical complex velocity given by $\de{w}/\de{z} = u - \i v = q\e^{-\i \theta}$. The two quantities on the left hand-side of \eqref{intanal} are complex-valued, but reduce to the physical speed and velocities on the free surface, $\zeta \geq 0$. Although their \emph{combined} values are related to the complex velocity, $u - \i v$, their individual values are not known without further work. 

As an example of this particular subtlety, we may consider the analytic continuation of $\theta$, evaluated along the physical boundary, $\zeta \leq 0$. In general, this value will not be the physical angle as defined in \eqref{steptheta} or \eqref{shiptheta}. Instead, the analytic continuations of the individual components, $q$ and $\theta$, are related to the physical angle through $\Im( q\e^{-\i\theta})$. 


In the bulk of this paper, we will focus on the analytic continuation into the upper half-plane, and thus set $k = 1$. Since $q$ and $\theta$ are real on the axis, then their values in the lower half-plane follows from Schwarz's reflection principal (see \S\ref{sec:hankel}). Then the two analytically continued governing equations are given by
\begin{subequations}
\begin{gather}
\ep q^2 \dd{q}{w} + \sin\theta = 0,  \label{bern_anal} \\
\log q + \im\j \theta = \log q_s + \j \bH[\theta]. \label{bdint_anal} 
\end{gather}
\end{subequations}

It is possible to combine the two equations into a single complex-valued integro-differential equation. We write write the sine term in terms of complex exponentials using \eqref{bdint_anal}, giving
\begin{equation} \label{sinsimp}
\sin\theta = \frac{1}{2\im}\left[ \left(\frac{q_s}{q}\right)^j \e^{\bH[\theta]} - \left(\frac{q}{q_s}\right)^j \e^{-\bH[\theta]}\right].
\end{equation}
We then simplify \eqref{sinsimp} using $\j = \pm 1$, and substitute into Bernoulli's equation \eqref{bern_anal} to obtain the combined integro-differential formulation.

\begin{gov*}[Analytic continuation of water waves \label{gov:analwaves}] We seek to study the analytically continued equations for the speed, $q$, and angle $\theta$, off the free surface and into the upper half-$\zeta$-plane. The combined integro-differential formulation is
\begin{equation} \label{combinedeq}
  \ep \dd{q}{w} + \frac{\im\j}{2} \frac{1}{q_s q^3}\biggl[ q^2 \e^{-\j\bH[\theta]}
-  q_s^2 \e^{\j\bH[\theta]} \biggr] = 0.
\end{equation}
which forms two equations (real and imaginary parts) for the two unknowns $q$ and $\theta$. In \eqref{combinedeq}, $q_s$ is contains the problem geometry as defined by \eqref{qs_step} and \eqref{qs_ship}, the analytically continued Hilbert transform, $\bH$, is defined in \eqref{Hrelate}, and we have defined the constants
\begin{equation} \label{jsign}
    \j = \begin{cases}
    1 & \text{for surface-piercing flow,} \\
    -1 & \text{for channel flow.}
    \end{cases}
\end{equation}
For the channel geometry, we use the conformal map from the potential strip to the $\zeta$-plane given by $\zeta = \e^{-w}$, while for the surface-piercing geometry, $\zeta = w$. 
\end{gov*}

Note that in all subsequent formulae, we will sometimes use $\zeta$ and $w$ interchangeably in the functional notation, \emph{e.g.} writing $q = q(w)$ or $q = q(\zeta)$ depending on the particular context. In general, primes ($'$) will be used solely for differentiation in $w$. 

\section{Connections to Tulin's formulation} \label{sec:tulinconnect}

The reader will notice that we have somewhat strayed from our original motivation of studying Tulin's formulation presented in \S\ref{sec:subtulin}, where the governing equations are posed in terms of a combined analytic quantity, $G(w)$, in \eqref{bern3}. Tulin's formulation can be adjusted to use the substitution $G = (q\e^{-\im\j \theta})^3$, where $\j = \pm 1$ as in \eqref{jsign}, which modifies the analytically continued Bernoulli's equation to 
\begin{equation}\label{tulinagain}
\frac{\ep}{G} \dd{G}{w} - \frac{\im\j}{G}  + \P(w, q, \theta)  = -\Q(w),
\end{equation}
for function, $\P$ with $\Re(\P) = 4 \sin^3\theta/q^3$, which had resulted from the sine reduction \eqref{sinreduce}, and for some $\Q$ that has yet to be specified. The adjustment of including $\j = \pm 1$ serves to orient the fluid with respect to the solid boundary, and facilitates comparison with our \eqref{combinedeq}.

We highlight two main differences between our \eqref{combinedeq} and Tulin's formulation \eqref{tulinagain}. First, in ours, we have chosen to work with analytic continuations of $q$ and $\theta$ independently, whereas Tulin has combined $G = (q\e^{-\i\j\theta})^3$, so as to advantageously write the left hand-side of \eqref{tulinagain} in an elegant form. However, note the nonlinear term, $\P$, cannot be easily written as a function of $G$, and this nonlinear term will be important in the analysis to come. It is unclear whether it is possible to provide a complete formulation without independentally treating $q$ and $\theta$, as we have done in our \eqref{combinedeq}. 

The second main difference is that \eqref{combinedeq} is self-contained and thus readily solved numerically or asymptotically, whereas the $\Q$-function in \eqref{bern3} is unknown. Like the introduction of $G$, while the introduction of the $\Q$-function is elegant in appearance, its connection to the low-Froude asymptotics is difficult to elucidate. 

We have included a discussion of Tulin's original interpretation of the $\Q$-function in Appendix~\ref{sec:Qfunc}. To summarize the conclusions of the discussion: the $\Q$-function can be written as a boundary integral evaluated over the solid boundary and its reflected image in the potential or $\zeta$-plane. However, this resultant quantity cannot be known without first solving the free-surface problem \eqref{bern} and \eqref{bdintgen}. Thus, $\Q$ should rather be written
\begin{equation}
  \Q(w) = \Q(w, q(w), \theta(w)),
\end{equation}
and it becomes difficult to unravel components of $\Q$ that are crucial to the determination of the free-surface waves. Our formulation \eqref{combinedeq} sidesteps this by explicitly including the local and global aspects of the problem in form that can be computed numerically and asymptotically. Once we have understood how \eqref{combinedeq} is reduced in the $\ep \to 0$, it will be possible to demonstrate more clearly how Tulin's formulation relates. We will do this in Appendix~\ref{sec:tulinconnect2}.

\section{Reduction to a simplified model} \label{sec:reduce}

The study of the nonlinear integro-differential equation \eqref{combinedeq} is primarily complicated because of the nonlocal nature of the Hilbert transform $\bH$. In \S\ref{sec:introtuck}, we had reviewed the efforts of E.O. Tuck, who attempted to convert $\bH$ into a local operator. In this section and the next, we shall demonstrate that, while the Hilbert transform is crucial for determining the corrections to the rigid-body solution (\emph{c.f.} the dotted curve in Figs.~\ref{fig:profilestern}), the waves can be derived largely independently from $\bH$.

We first introduce the truncated regular expansion
\begin{subequations} \label{qthetasub}
\begin{equation} \label{qrtrser}
q_r = \sum_{n=0}^{N-1} \ep^n q_n  
\qquad \text{and} \qquad
\theta_r = \sum_{n=0}^{N-1} \ep^n \theta_n, 
\end{equation}
where $q_r$ and $\theta_r$ are seen to be waveless---that is to say, they do not possess any oscillatory component. The oscillatory components will be introduced via a linearization of the solution about the regular expansion. We set,
\begin{equation} \label{qrtr}
	q = q_r + \bq \qquad \text{and} \qquad
	\theta = \theta_r + \bt,
\end{equation}
\end{subequations}
where we assume $\bq, \bt = o(\ep^{N-1})$.

Our strategy is to substitute \eqref{qthetasub} into the integro-differential equation \eqref{combinedeq}, and then separate the terms proportional to $\bq$ onto the left hand-side. Simplification then leads to the main result of this paper.

\begin{result}[Reduced integro-differential model] \label{result:integro}
Linearizing the water wave equations about a regular series expansion truncated at $N$ terms gives the following integro-differential equation for the perturbation, 
\begin{subequations} \label{simpsys}
\begin{equation} \label{simpsyseq}
\ep \bq' + \biggl[ \chi'(w) + \ep P_1'(w) + \Oh(\ep^2) \biggr]\bq 
= R(w; \bH[\bt]) + \Oh(\bt^2, \bq^2).	
\end{equation}
where
\begin{gather}
\chi'(w) = \frac{\im \j}{q_0^3(w)}, \label{chip} \\
P_1'(w) =  \left(2 \frac{q_0'}{q_0} - 3(\im \j)\frac{q_1}{q_0^4}\right), \label{P1p} \\
R(w; \bH[\bt]) = -\Eb + \im \bH[\bt]\frac{\cos\theta_r}{q_r^2}.	\label{Rfunc}
\end{gather}
and $q_0$ and $q_1$ are given in \eqref{asym0} and \eqref{asym1}, and the error term, $\Eb$, represents the error in Bernoulli's equation, and is given by
\begin{equation} \label{Ebern}
  \Eb = \ep q_r' + \frac{\sin\theta_r}{q_r^2}.
\end{equation}
\end{subequations}
\end{result}

Note that the function $\chi$ that appears in \eqref{chip} is precisely the same as the singulant function that describes the factorial-over-power divergence \eqref{qt_fact} and whose value is found in Appendix~\ref{sec:divergence} through the study of the regular asymptotic expansion of the solutions. Similarly, in the next section, $P_1$ will be related to $Q$, which appears in \eqref{qt_fact}.

We will use the notation of $R(w)$ and $R(w; \bH[\bt])$ interchangeably, preferring the latter when we wish to emphasize that $R$ depends on the inclusion of the Hilbert transform. Notice also that the system \eqref{simpsys} will undergo significant changes in its written form as soon as $w = \phi + \im \psi$ or $\zeta = \xi + \im \eta$ approaches the free surface (either $w \in \mathbb{R}$ or $\zeta \in \mathbb{R}^+$). When this occurs, the integral $\bH[\bt]$ must be written as a principal value and residue contribution, as in \eqref{hilbertsplit}. In other words, though it is written in a complex-valued form, \eqref{simpsys} will necessarily reduce to a real-valued equation along the free surface. 

\subsection{Truncation and optimal truncation} \label{sec:truncation}

The crucial decision in studying the system \eqref{simpsys} is to choose how many terms to include in the truncated series, $q_r$ and $\theta_r$, which determines the forcing function $R$. That is, we must choose the truncation value of $N$. 

In early studies of the low-Froude problem, researchers had used $N = 1$, and then later corrected to $N = 2$. However, at a fixed value of $\ep$, the truncation error in the divergent asymptotic expansion will continue to decrease for increasing values of $N$. It reaches a minimum at an optimal truncation point, $N = \mathcal{N}$, and diverges afterwards. For small $\ep$, the optimal truncation point is found where adjacent terms in the series are approximately equal in size, or $| \ep^\N q_\N | \sim | \ep^{\N-1}q_{\N-1}|$. Using the divergent form \eqref{qt_fact}, we find that the optimal truncation point is given by 
\begin{equation}
  \N \sim \frac{|\chi(w)|}{\ep}.
\end{equation} 
Thus, for a fixed point in the domain, $w$, and in the limit $\ep \to 0$, the optimal truncation point, $\N$, tends to infinity, and we must include infinitely many terms of the series. In \S\ref{sec:numerics}, we will examine the effect of different truncations on the comparison between numerical solutions and asymptotic approximations.

Most emphatically, we remark that apart from the hidden $\Oh(\ep^2)$ terms multiplying $\bq$ in the brackets of \eqref{simpsyseq}, this reduced equation is an exact result to all orders in $\ep$. We have only employed the fact $\bq, \bt \ll 1$, but the inhomogeneous term, $R(w)$ is exact. When the regular series expansions are optimally truncated, $\bq$ and $\bt$ are exponentially small, and consequently, the $\Oh(\bq^2, \bt^2)$ will not be needed to derive the leading-order behaviour of the exponentials. 

Though it is now written as a first-order differential equation, the system \eqref{simpsys} is difficult to solve, since it involves real and complex components, in addition to the integral transforms embedded in the inhomogeneous term. The key in the following sections will be to argue that the Hilbert transform can be neglected. 



\section{General analysis using steepest descents} \label{sec:steep}

The main goal in this section is to demonstrate how the reduced integro-differential equation \eqref{simpsys} can be written as an explicit integral, which is then approximated using the method of steepest descents (for details of this method, see for example \citealt{bleistein_1975_asymptotic_expansions}). It is, in fact, this steepest descents analysis that rectifies the unanswered issues from Tulin's work. 

In this paper, we will present the main ideas of a steepest descents analysis applied to the integral solution of \eqref{simpsys}. In practice, the individualized steepest descent paths must be studied for each different moving body (as specified by the $q_s$ function in \eqref{logqs}). In a companion paper \citep{trinh_tulinsteep_paper}, we will derive the full steepest descents structure for the case of flow over a step and past the stern of a ship. This individualized study requires careful consideration of the branch structures and numerically generated contours of the integrand, so we relegate the details to the companion study.

\subsection{Integral form of the solution} 

To begin, we integrate \eqref{chip} to obtain
\begin{equation} \label{chi}
\chi(w) = \im \j \int_{w_0}^w \frac{\de{\varphi}}{q_0^3(\varphi)}.
\end{equation}
In the above integral, the initial point of integration, $w_0$, is chosen to be the particular point that causes the divergence of the late-order terms \eqref{qt_fact}. Thus, the $\chi$ defined in \eqref{simpsys} matches the definition of the singulant function in the late-orders ansatz. 

Next integrating \eqref{P1p},
\begin{equation} \label{P1}  
P_1(w) =  \e^{-\Lambda} + \log \left[\frac{q_0^2(w)}{q_0^2(w^*)}\right] - 3\im \j \int_{w^*}^w \frac{q_1(\varphi)}{q_0^4(\varphi)} \, \de{\varphi}, 
\end{equation}
where $\e^{-\Lambda}$ is the constant of integration, and $w^*$ can be chosen wherever the integral is defined. We also note that $P_1$, given by \eqref{P1}, is related to $Q$ given in \eqref{Q}. Thus we can write 
\begin{equation} \label{expP1}
\e^{-P_1(w)} = q_0^2(w^*) Q(w) = \left[\frac{\Lambda q_0^2(w^*)}{q_0^2(w)}\right] \exp \left( 3\im j k \int_{w^*}^w \frac{q_1(\varphi)}{q_0^4(\varphi)} \, \de{\varphi}\right).
\end{equation}

Solving \eqref{simpsyseq} now yields a solution for the integro-differential equation (though note that it is not quite explicit due to the reliance on the Hilbert transform),
\begin{subequations} \label{finalint}
\begin{equation} \label{finalq}
\bq(w) = \left[\frac{Q(w)}{\ep}\right]\biggl\{ I(w)  + \text{const.} \biggr\} \times \e^{-\chi(w)/\ep} 
\end{equation}
where we have introduced the integral 
\begin{equation} \label{Iint}
I(w) = \int_{w_s}^w R(\varphi; \bH[\bt]) \left[ \frac{1}{Q(\varphi)} + \Oh(\ep)\right] \, \e^{\chi(\varphi)/\ep}\, \de{\varphi}.	
\end{equation}
\end{subequations}
which will be used to extract the pre-factor of the exponential. The start point, $w_s$, can be conveniently chosen based on any additional boundary or radiation conditions. For channel flow, the water wave problem will naturally impose a radiation condition which requires the free surface to be flat at $\phi = -\infty$ (upstream), so taking $w_s \to -\infty$, the constant of integration in \eqref{finalq} is zero. Similarly, for the surface-piercing problem, $w_s$ can be taken to be the stagnation point attachment, $w_s = 0$, where $q = 0$. 

\subsection{The steepest descent paths} 

\begin{figure}\centering
\vspace*{1.0\baselineskip}
\includegraphics[width=1.0\textwidth]{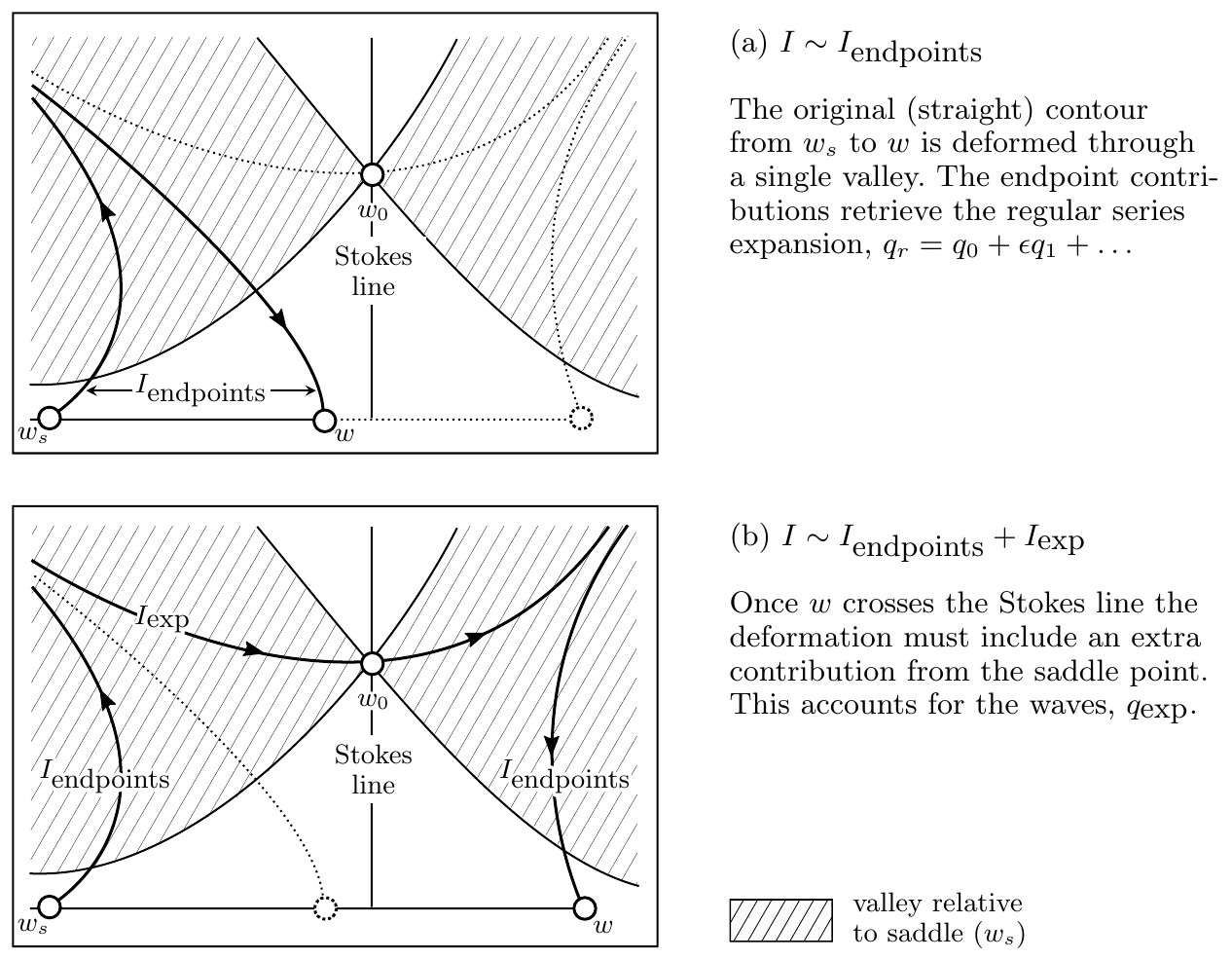}
\caption{Illustration of steepest descent contours for a problem with a single wave-generating singularity, $w = w_0$. (a) When the endpoint of integration lies to the left of the Stokes line, $I \sim I_\text{endpoints}$; (b) when the end of integration crosses the Stokes line, the steepest descent path must include the contribution from the saddle. \label{fig:steepest}}
\end{figure}

For the integral \eqref{Iint}, the paths of steepest descent are given by constant contours of $\Im \chi$. Let us envision a generic situation in which the steepest descent topology is as shown in the Fig.~\ref{fig:steepest}. The valleys, where $\Re[\chi(w)] \geq \Re[\chi(w_0)]$, are shown shaded. In the illustrated case, the path of steepest descent from the initial, $t = w_s$,  and final point, $t = w$, is joined by a single valley of the integrand. Thus, we write
\begin{equation} \label{Iend}
  I(w) \sim I_\text{endpoints},
\end{equation}
and the relevant contributions to the integral solely depend upon an expansion about the endpoints. When we perform this analytically in \S\ref{sec:endpoint}, we will find that the $I_\text{endpoint}$ contribution will yield further algebraic orders of the base asymptotic series, $q_r$ and $\theta_r$. In other words, contributions \eqref{Iend} lead to further deflections of the free surface due to the moving body, but do not produce an oscillatory component.

In the free-surface problems we consider, the oscillatory contributions arise from the deformation of the path of integration through any critical points (branch points, singularities, or saddle points) of the integrand or exponential. A typical situation is illustrated in Fig.~\ref{fig:steepest}(right). There, we see that if the endpoint, $t = w$, crosses the Stokes line, $\Im[\chi] = \Im[\chi(w_0)]$ from the point, $t = w_0$, then the steepest descent paths from each individual endpoint tends to distinct valleys that must be joined by a contour through the saddle. In such cases, we shall write
\begin{equation} \label{Iboth}
  I(w) \sim I_\text{endpoints} + I_\text{exp}.
\end{equation}

In this work, we shall not discuss the global properties of the steepest descent curves and and the associated Stokes lines. These results require careful consideration of the Riemann-sheet structure of the integrand, and can be found in the accompanying work \citep{trinh_tulinsteep_paper}. Instead, we will assume that the integral is approximated by \eqref{Iboth}, and verify the approximation for the case of the far-field waves, $w \to \infty$, in \S\ref{sec:numerics}.

\subsection{Endpoint contributions} \label{sec:endpoint}

It can be verified that the dominant contributions from the endpoints will re-expand the regular perturbation series to higher orders. In particular, we see from inspection that locally near $t = w$, the integrand is exponentially large and of order $\e^{\chi/\ep}$, and this serves to cancel the exponentially small factor in \eqref{finalq}. 

The form of the integral in \eqref{finalint} is unwieldy, so we shall demonstrate this re-expansion for the simplest case where the regular series, $q_r$ and $\theta_r$, contains only a single term. Setting $q_r = q_0 = q_s$ and $\theta_r = \theta_0 = 0$ from \eqref{asym0}, the integral is reduced to
\begin{equation}
I_\text{endpoints} \sim \int_{w_s}^w \biggl[ -\ep q_0' + \frac{\im \bH[\bt] }{q_0^2} \biggr] \left[ \frac{1}{Q(\varphi)} + \Oh(\ep)\right] \, \e^{\chi(\varphi)/\ep} \, \de{\varphi}.
\end{equation}
Integrating by parts, we have
\begin{multline}
I_\text{endpoints} \sim \frac{\ep}{\chi'} \biggl[ -\ep q_0' + \frac{\im \bH[\bt] }{q_0^2} \biggr] 
\left[ \frac{1}{Q(w)} + \Oh(\ep)\right] \, \e^{\chi(w)/\ep} \\ 
- \ep \int_{w_s}^w \frac{1}{\chi'} \dd{}{\varphi} 
\left\{\biggl[ -\ep q_0' + \frac{\im \bH[\bt] }{q_0^2} \biggr] \left[ \frac{1}{Q(\varphi)} + \Oh(\ep)\right] \right\} \, \e^{\chi(\varphi)/\ep} \, \de{\varphi},
\end{multline}
where we have assumed that the boundary term contributions due to $\varphi = w_s$ are zero from the boundary or radiation conditions (see the discussion following \eqref{Iint}). Repeated integration by parts to the integral term will extract further contributions. We keep only the first term above, and re-combine with $\bq$ in \eqref{finalq} to obtain
\begin{equation}
	\bq \sim \frac{q_0^3}{\im \j} \biggl[ -\ep q_0' + \frac{\im \H[\bt]}{q_0^2} - \frac{\bt}{q_0^2}\biggr]. 
\end{equation}
where we have used \eqref{Hrelate} and \eqref{chip}. Matching real and complex parts then yields $\bq = \j q_0 \H[\bt]$ and $\bt = -\ep q_0^2 q_0'$, which are the first correction terms derived in \eqref{asym1}. The procedure to obtain further corrections proceeds in a similar way using repeated integration by parts, and thus producing further powers of $\ep$. 

Including more terms from $q_r$ and $\theta_r$ and performing the re-expansion procedure about the endpoints will follow the same idea; algebraically, however, it is much easier to work with the integro-differential equation \eqref{simpsyseq}, or simply with the original system of two coupled equations, as was done in \S\ref{sec:baseasym}.

\subsection{The Hankel contour} \label{sec:hankel}

In the limit $\ep \to 0$, the contributions to the contour integral associated with wave motion arises from cases where the contour is deformed past critical points in the integrand, or near saddle points, where $\chi'(t) = 0$. Let us assume that the contribution occurs at a point, $w = w_0$, where $\chi'(w_0) = 0$. We furthermore assume that $w_0$ lies off the free surface (see \S\ref{sec:tulinlowspeed} for comments on other cases). Locally near this point, we assume that the leading-order speed behaves as
\begin{equation} \label{nearq0}
	q_0 \sim c (w - w_0)^\alpha,
\end{equation}
for some constants $c$ and $\alpha$, and consequently, it follows from \eqref{chip} that
\begin{equation} \label{chiX}
	\chi \sim X (w - w_0)^{1 - 3\alpha} \qquad \text{where} \quad 
	X = \frac{\im \j}{c^3(1-3\alpha)}.
\end{equation}

One of the key issues that underlies the Tulin reduction of the water-wave equations concerns the role of the non-local Hilbert Transform, $\bH$, which appears in the forcing function $R(w; \bH[\bt])$ in \eqref{Rfunc}. In the previous section, we discovered that the Hilbert Transform plays an important role in the further development of the endpoint contributions. Indeed, it appears as a contributing term in each order of the asymptotic process, beginning from $q_1$ in \eqref{asym1}. 

Upon applying the method of steepest descents to the integral $I$, we had separated the contributions into those due to endpoints and those due to saddle points (assumed to lie away from the free surface). Let us assume that the contribution due to the saddle point is exponentially small along the free surface, so that $\bq_{\text{exp}} = \Oh(\e^{-\chi/\ep})$ in \eqref{finalq}, with $\Re(\chi) > 0$ on the free surface, and similar relations for $\bt$. Then $I_\text{exp}$ is at most algebraically large in $\ep$ and near the singularity, $t = w_0$, and the term
\begin{equation} \label{hilbertargue}
	\im \bH[\bt] \frac{\cos \theta_r}{q_r^2}
\end{equation}
involves the integration of an exponentially small term along the free surface, where it remains exponentially small. Thus, this term is subdominant to the square-bracketed contributions in \eqref{Rfunc}. Further, note that this argument does not work for the endpoint contributions, as the Hilbert transform within $\Ei$ is \emph{not} negligible, as it involves the integration of the regular perturbative series via $\bH[\theta_r]$. 

Let us first consider the case of truncation at large $N$. As discussed in \S\ref{sec:truncation}, it is expected that as $\ep \to 0$, the optimal truncation point, $\N \to \infty$. Thus using the divergent form \eqref{qt_fact}, we have that away from the real axis,
\begin{equation} \label{nearR}
  R \sim  -\ep^N q_{N-1}' \sim \ep^N \frac{Q\Gamma(N+\gamma) \chi'}{\chi^{N+\gamma}},
\end{equation}
where recall that the components, $Q$, $\gamma$, and $\chi$ are given in Appendix~\ref{sec:divergence}. Thus substituting \eqref{nearR} into \eqref{Iint}, we have
\begin{equation} \label{Iexpa}
  \Iexp \sim \ep^N \Gamma(N+\gamma) \int_{C_0} \left[\frac{\chi'}{\chi^{N+\gamma}}\right] \e^{\chi/\ep} \, \de{\varphi}.
\end{equation}
In \eqref{Iexpa}, the contour $C_0$ corresponds to the steepest descents contour in the vicinity of the critical point, $\varphi = w_0$. Near this point, we simplify the exponential by making a coordinate transformation with
\begin{equation} \label{usub}
  u = \frac{X(\varphi - w_0)^{1-3\alpha}}{\ep} \quad \text{and thus} \quad \chi \sim \ep u
\end{equation}
from \eqref{chiX}. Using \eqref{usub}, we see that $\dd{\chi}{\varphi} \de{\varphi} \sim \ep \de{u}$, and then 
\begin{equation} \label{Iexp_u}
  \Iexp \sim
  \frac{\Gamma(N+\gamma)}{\ep^{\gamma - 1}} \int_{C_0} u^{-(N+\gamma)} \e^{u} \, \de{u} 
\end{equation}

We claim from the steepest descent topographies that for the ship and step problems, the relevant local contribution near $\varphi = w_0$ occurs in the form of a Hankel contour---that is, an integral about the branch cut $u \in \mathbb{R}^-$, beginning from $u = -\infty - 0\im$, looping around the origin, and tending to $u = -\infty + 0\im$. These steepest descent topographies are the subject of the companion paper \cite{trinh_tulinsteep_paper}. By the integral definition of the Gamma function,
\begin{equation} \label{hankel}
	\cint_{C_0} u^{-(N+\gamma)} \mathrm{e}^u \, \de{u} = \frac{2\pi \im}{\Gamma(N + \gamma)}.
\end{equation}
Therefore, combining with \eqref{Iexp_u}, we have $\Iexp \sim 2\pi \im \ep^{1-\gamma}$. We may now substitute this approximation for $\Iexp$ into \eqref{finalq} to obtain the part of $\bq$ that is switched-on by the saddle point contribution. This yields
\begin{equation} \label{bqexp}
  \bqexp \sim \left[\frac{2\pi \im Q(w)}{\ep^\gamma}\right] \e^{-\chi(w)/\ep},
\end{equation}
where $Q(w)$ is given in \eqref{expP1}. 

However, we recall that the functional form of integrand in \eqref{Iint} changes in the limit the real axis is approached, due to the presence of the complex Hilbert transform in $R(\varphi)$. Moreover, the derivation of \eqref{bqexp} takes account the steepest descent paths only for analytic continuation into either the upper-half $w$-plane ($\k = 1$) or lower-half $w$-plane ($\k = -1$). Repeating the process for the opposite portion of the plane reflects the steepest descent paths about the real axis.

Rather than repeating the procedure and observing the change in signs of $\k$, we can note that wave component of $q$ or $\theta$ must be real along the physical free surface. Consequently, by the Schwarz reflection principle, the leading-order wave solution must be the sum of \eqref{bqexp} and its complex conjugate. Thus on the free surface, where $w = \phi$,
\begin{equation} \label{qexpcc}
	\qexp(\phi) \sim \left[\frac{2\pi \im Q(\phi)}{\ep^\gamma}\right] \e^{-\chi(\phi)/\ep} + \textrm{complex conjugate.}
\end{equation}


Note that in numerical computations, it is typically easiest to compare the wave amplitudes in the far field, where $w = \phi \to \infty$, and the speed, $q_r \to 1$. Substituting \eqref{expP1} into \eqref{qexpcc} and simplifying gives the amplitude of the oscillations,
\begin{equation} \label{amplitudeq}
\text{Amplitude of waves in $q$} = \left[\frac{2\pi|\Lambda| \exp\left(\im\j \int_{w^*}^w \frac{q_1}{q_0^4} \, \de{\varphi}\right)}{\ep^\gamma}\right] \e^{-\Re(\chi)/\ep}, \quad \text{as $\phi \to \infty$},
\end{equation}
where the constant $\Lambda$ appears in the expression for $Q$ and its value must be computed numerically for general nonlinear problems (see Appendix~\ref{sec:divergence}). Although the above formula seems to depend on the integral limit $w^*$ (assumed to be chosen anywhere the integral is defined), this is in fact compensated by the calculation of $\Lambda$. In the end, \eqref{amplitudeq} provides a closed-form formula for the amplitude of the waves, which only depends on specification of the two solutions, $q_0$ and $q_1$, calculated from \eqref{asym01}, and late-order components, $\Lambda$, $\gamma$, and $\chi$, given in Appendix~\ref{sec:divergence}. We will provide a numerical example of \eqref{amplitudeq} shortly.

Finally, we can establish an equation between the the exponentially small waves in $q$ and $\theta$. From \eqref{bdint_anal}, upon substituting \eqref{qthetasub}, we have the connection, 
\begin{equation} \label{theta-qb}
  \texp = \left[\frac{\i \j}{q_0} + \Oh(\ep)\right]\qexp,
\end{equation}
which allows the amplitude of the waves to be calculated, both in $q$, as given by \eqref{amplitudeq}, and in $\theta$. 

\section{Comparisons of the full nonlinear model with the reduced models} \label{sec:numerics}

In the last two sections, we derived a reduced model for the wave-body interaction, and subsequently explained how asymptotic approximations for both the wavefree surface deflection (the regular series $q_r$) and the waves (the exponentially small $q_\textrm{exp}$) could be developed using the method of steepest descents. Now the question of how this model relates to both the \cite{tulin_1982_an_exact} and the \cite{tuck_1990_water_non-waves, tuck_1991_ship-hydrodynamic_free-surface, tuck_1991_waveless_solutions}, along with the other possible simplifications discussed in \S\ref{sec:otherworks}, will depend on how we decide to truncate the model in \eqref{simpsys}. 

\subsection{Truncation at other values of $N$} \label{sec:steeptrun}

The result of the Hankel integration, in \eqref{qexpcc}, which depends on the late-terms behaviour of the ansatz \eqref{qt_fact}, was derived on the assumption that truncation of the regular series expansion occurs at optimal truncation with $N \to \infty$ as $\ep \to 0$. In practice, however, it may be useful to develop reduced models that truncate at pragmatic values of $N$ (\emph{i.e.} one or two); let us discuss what loss in accuracy results from this.

As we have shown, the surface speed is expressed as
\begin{equation}
  q(\phi) \sim \Bigl[q_0(\phi) + \ep q_1(\phi) + \ldots + \ep^{N-1} q_{N-1}(\phi)\Bigr] + \Bigl[\mathcal{A} F(\phi) \e^{-\chi(\phi)/\ep} + \text{c.c.}\Bigr],
\end{equation}
where we have expressed the exponential \eqref{bqexp} in a form so as to separate a numerical prefactor, $\mathcal{A}$, from the functional dependence, $F(\phi)$. 



The case of truncation at $N = 1$ is asymptotically inconsistent since if $q = q_0 + \bq$, then $\bq$ contains both an $\Oh(\ep)$ error and also the exponentially small waves we desire. Thus from \eqref{simpsyseq} and \eqref{P1p}, we see that it would not be correct to ignore $\Oh(\bq^2)$ errors in deriving the linearized equation. Despite these issues, however, it is still instructive to examine the truncation at $N =1$, which yields 
\begin{equation} \label{N1eq}
\ep \bq' + \chi' \bq \sim -\ep q_0,
\end{equation}
and thus the solution
\begin{equation}
  \bq \sim - \left[ \int_{w_s}^w q_0 \e^{\chi/\ep} \, \de{t} \right]\e^{-\chi/\ep},
\end{equation}
for a start point, $t = w_s$, where $\bq(w_s) \to 0$. 

We then see that the exponential argument, $\e^{-\chi/\ep}$, of the solution could be derived from \eqref{N1eq}, and indeed the steepest descent paths of \S\ref{sec:steep} are still applicable, but we have inaccurately predicted the functional dependence, $F(\phi)$ (captured by the missing $O(\ep \bq)$ term on the left hand-side), and will also fail to predict the prefactor, $\mathcal{A}$ (captured by the correct right hand-side). 

The case of truncation at $N = 2$ is much more interesting. It yields
\begin{equation} \label{qbN2}
\ep \bq' + \biggl[ \chi'(w) + \ep P_1'(w)\biggr]\bq 
\sim \ep^2 \left(-\frac{5\im\j q_1^2}{2 q_0^4} + q_1' + 2\im \frac{\bH[\theta_1]q_1}{q_0^3}\right).
\end{equation}
Thus, comparing \eqref{qbN2} to \eqref{simpsys}, we have obtained the precise left hand-side required. In developing the leading-order exponential, the only error in \eqref{qbN2} is due to the replacement of the right hand-side by its $N = 2$ truncation. This only affects the integrand of \eqref{finalint}, whose role was to determine the pre-factor $\mathcal{A}$. In other words, the linear equation \eqref{qbN2} allows us to develop \emph{nearly all} of the leading-order exponential. We will see in the numerical computations that this equation provides a very close fit to results from the full nonlinear model. The two-term truncation is the closest analogy to the \cite{tulin_1982_an_exact} model (see \S\ref{sec:discuss} for an extended discussion), as well as various other models (\emph{c.f.} the review by \citealt{doctors_1980_comparison_of}). However the fact that such two-term approximations do not accurately predict the pre-factor $\mathcal{A}$ is not well established in the literature.

\subsection{A simplified nonlinear model for $N = 2$}

An even simpler formulation to \eqref{qbN2} can be developed at the risk of slightly increased inaccuracy. However, this form serves as an extremely useful toy model due to the complete removal of the Hilbert transform---in this sense, it is the strongest analogy to the Tuck reduction of \eqref{tucksimp}. Let us return to combined integro-differential equation in \eqref{combinedeq}. Approximating $e^{\j\bH[\theta]} \sim 1$, we obtain
\begin{equation} \label{simpnonlinear}
  \ep q_s q^3 q' + \frac{\im \j}{2} \left[ q^2 - q_s^2 \right] = 0.
\end{equation}

In order to avoid confusion, we shall write the solution of \eqref{simpnonlinear} as $\tilde{q}$. Expressing $\tilde{q}$ as a regular series expansion $\tilde{q} = \tilde{q}_0 + \ep \tilde{q}_1 + \ldots$, we find 
\begin{equation}
 \tilde{q_0} = q_s \quad \text{and} \quad \tilde{q}_1 = \im\j q_0^3 q_0'.
\end{equation}
The leading-order solution is as expected, but ignoring the $\bH[\theta]$ complex Hilbert transform has the effect of changing the correction term, $\tilde{q}_1$. If we return to the correct $q_1$ in \eqref{asym1}, which corresponds to the full model, we find
\begin{equation}
  q_1 = \j q_0 \Bigl(\bH[\theta_1] - \im \theta_1\Bigr) = \im j q_0 q_0' + \j q_0 \bH[\theta_1]
\end{equation}
upon using \eqref{Hrelate} and the solution for $\theta_1$ in \eqref{asym1}. Recall that near the wave-generating singularity, $\bH[\theta_1]$ remains bounded, but $q_0$ and its derivatives are singular according to \eqref{nearq0}. In other words, the simplified formulation of \eqref{simpnonlinear} has replaced $q_1$ in the full model with its local behaviour near the singularity. 

Substitution of $q = \tilde{q}_0 + \ep \tilde{q_1} + \ldots + \bq$ into the simplified nonlinear model \eqref{simpnonlinear} yields
\begin{equation} \label{simpnonlinearlin}
\ep \bq' + \biggl[ \chi' + \ep \frac{5 q_0'}{q_0}\biggr]\bq 
\sim \tilde{R}(w),
\end{equation}
where we have withheld the right hand-side, $\tilde{R}$ for clarity. We can verify that as $w \to w_0$, $P_1' \sim 5 q_0'/q_0$, and thus comparing the bracketed terms in \eqref{qbN2} and \eqref{simpnonlinearlin}, we see that while we have completely neglected the $\bH[\theta]$ terms, we are nevertheless able to preserve the inner limit of \eqref{qbN2}. Thus, since the limiting behaviours of the $\bq$ coefficient functions are preserved exactly, then the Hankel contour analysis of \S\ref{sec:hankel} (which depends only on local properties) will also be preserved exactly, with the exception of a different numerical prefactor due to the right hand-side differences. 

In summary, let us assume that the leading-order exponential for the full problem is written as $\mathcal{A}F(\phi)\e^{-\chi/\ep}$ with limiting behaviour $F(\phi) \sim F_0(\phi - \phi_0)^\mu$ as $\phi \to w_0$, and for constants $F_0$ and $\mu$. Then the exponential that results from using the simplified nonlinear formulation of \eqref{simpnonlinear} is $(\phi - \phi_0)^\mu \tilde{A} \e^{-\chi/\ep}$ for some constant $\tilde{A}$. In the study of \cite{trinh_2015_exponential_asymptotics}, the simplified nonlinear problem \eqref{simpnonlinear} was used as a toy model for the study of wave-structure interactions with coalescing singularities. By duplicating the derivation of \S\ref{sec:steep}, the analogous formula to \eqref{amplitudeq} can be developed for \eqref{simpnonlinear}. We will provide a numerical computation of this problem in \S\ref{sec:numerics}. 

A detailed summary of the full and simplified models we have presented thus far is shown in Table~\ref{tab:models}.


\afterpage{
\clearpage
\begin{landscape}
\topskip0pt
\vspace*{\fill}
\begin{table} \centering
\begin{tabular}{cccccccp{8cm}} 
\scshape{truncation} 
& \scshape{equation}
& \qquad & $\chi$ & $F(\phi)$ & $\mathcal{A}$ \
& \qquad & \scshape{notes} \\[0.5\baselineskip]
none & $\displaystyle \ep q' +  \frac{\im\j}{2 q_s q_0^3} \Bigl(q^2 \e^{-\j\bH[\j\theta]} - q^2 \e^{-\j\bH[\theta]}\Bigr) = 0$ 
& \qquad & yes & yes & yes 
& \qquad & The full nonlinear problem \eqref{combinedeq} requires solving a nonlocal integro-differential formulation. Real and complex parts yield equations for $q$ and $\theta$. \\[0.5\baselineskip]
none & $\displaystyle \ep \frac{G'}{G} - \frac{\im\j}{G}  + \P(w, q, \theta)  = -\Q(w, q, \theta)$
& \qquad & yes & yes & yes 
& \qquad & Tulin's formulation \eqref{tulinagain}, in terms of $G = (q\e^{-\im\j\theta})^3$, is difficult to use due to unknown $\Q$. Note that nonlinear $\P$, with $\Re(\P) = 4\sin^3\theta/q^3$ cannot be written in terms of $G$, and neglecting $\P$ is equivalent to a truncation at $N = 2$. \\[0.5\baselineskip]
$N = 1$ & $\displaystyle \ep \bq' + \chi' \bq = \Oh(\ep)$ 
& \qquad & yes & no & no 
& \qquad & From \eqref{N1eq}; despite simplicity, contains many key properties of the water waves; one-term truncation used in \cite{ogilvie_1968_wave_resistance:}. \\[0.5\baselineskip]
$N = 2$ & $\ep \bq' + (\chi' + \ep P_1')\bq = \Oh(\ep^2)$
& \qquad & yes & yes & no 
& \qquad & From \eqref{qbN2}; the simplest model that preserves everything but the prefactor $\mathcal{A}$. Provides a very close fit to the full nonlinear solutions. \\[0.5\baselineskip]
$\phantom{^*}N = 2^*$ & $\ep q_s q^3 q' + (\im\j/2)(q^2 - q_s^2) = 0$ 
& \qquad & yes & no$^*$ & no
& \qquad & From \eqref{simpnonlinear}; simplified nonlinear model that uses inner limit to replace the Hilbert transform; used as toy model in \cite{trinh_2015_exponential_asymptotics}.\\[0.5\baselineskip]
$N \to \infty$ & $\ep \bq' + (\chi' + \ep P_1')\bq = \Oh(\ep^{N})$ 
& \qquad & yes & yes & yes 
& \qquad & From \eqref{simpsyseq} with \eqref{nearR}; the \emph{correct} linearized equation that preserves all aspects of the leading-order exponential; requires optimal truncation and numerical computation of the prefactor $\mathcal{A}$. \\
\end{tabular}
\caption{A comparison of different truncated models for the study of gravity waves past moving bodies, and their ability to correctly predict (yes/no) the leading-order exponential, $\mathcal{A} F(\phi)\e^{-\chi/\ep}$ (to be added to its complex conjugate on the free surface). The case of $N = 2^*$ is a two-term truncation that extracts the singular behaviour of the Hilbert transform, and thus obtains the correct functional form of $F(\phi)$ near the relevant singularity, $w = w_0$. \label{tab:models} }
\end{table}
\vspace*{\fill}
\end{landscape}
}

\subsection{Numerical comparisons with the full water wave equations}

In this section, we will verify the fit of the three reduced models with the numerical solutions of the full nonlinear problem in the limit $\ep \to 0$. The models include (i) the full nonlinear problem \eqref{combinedeq}, or more conventionally, the solution of Bernoulli's equation \eqref{bern} and the boundary integral equation \eqref{bdintgen}; (ii) the $N = 2$ truncated linear model in \eqref{qbN2}; and (iii) the $N = 2^*$ nonlinear model in \eqref{simpnonlinear}. 

We study the case of a semi-infinite rectangular ship given by \eqref{shiptheta} and \eqref{qs_ship} with $\sigma = 1/2$, and thus
\begin{equation} \label{q0shipspec}
  q_0 = \left(\frac{w}{w+1}\right)^{1/2},
\end{equation}
which is the leading-order speed associated with a hull with a right-angled corner ($3\pi/2$ in the interior of the fluid) at $w = -1$, and a stagnation point at $w = 0$.

The solution is computed in each of the three cases, and the amplitude of the water waves far downstream, with $w = \phi \gg 1$, is extracted. For the two models with truncations at $N = 2$ and $N = 2^*$, recall that the real-valued solution on the axis is formed by adding the complex-valued complex solution to its complex conjugate [see \eqref{qexpcc} and the surrounding discussion]. Thus, for these two cases, the amplitude of $q$ is taken to be twice the amplitude of $\Re(\qexp)$.

\begin{figure} \centering
\includegraphics{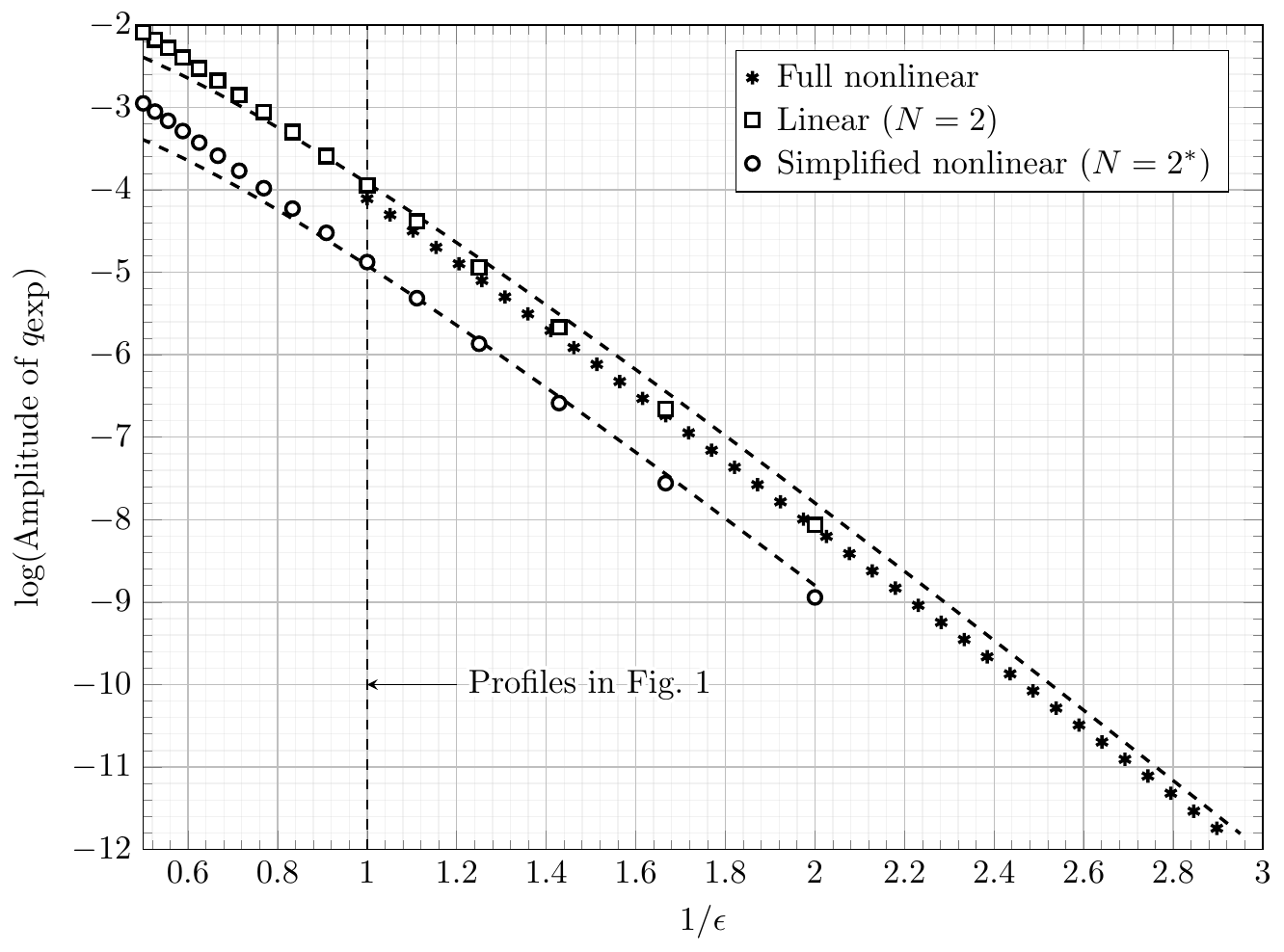}
\caption{Comparison of wave amplitudes for the full nonlinear model, the $N = 2$ truncated linear model, and the $N = 2^*$ truncated nonlinear model. For the two truncated models, the amplitude is multiplied by two to account for the analytic continuation. The dashed lines correspond to leading-order asymptotic approximations of the full nonlinear model (top) and the simplified nonlinear model (bottom). \label{fig:numerics}}
\end{figure}

In order to solve the full water wave equations \eqref{combinedeq}, we use the numerical algorithm described in \cite{trinh_2011_do_waveless}. In brief, a stretched grid is applied near the stagnation point, and a finite-difference approximation of the boundary integral is calculated using the trapezoid rule. At a singular point of the integral, a quadratic interpolant is applied between the point and its two neighbours, and the resultant quadratic is calculated exactly. For more details of the numerical scheme see, \emph{e.g.} Chap.~7 of \cite{vb_book} and the references therein. From the predicted wave amplitude \eqref{amplitudeq}, we have
\begin{equation} \label{ship12amp}
  \text{Amplitude of $\qexp$} = \left[\frac{\tilde{C}}{\ep^\gamma}\right] \e^{-3\pi/(2\ep)},
\end{equation}
where $\tilde{C} \approx 2.215$, $\gamma = 6\sigma/(1+3\sigma)$, and here, $\sigma = 1/2$. The numerical pre-factor $\tilde{C}$ requires a generic calculation (\emph{c.f.}  Fig.~10 of \citealt{trinh_2011_do_waveless}). Both numerical amplitude measurements (stars), as well as the asymptotic prediction (upper dashed line), are shown in Fig.~\ref{fig:numerics}.

The $N = 2$ and $N = 2^*$ truncated models can be solved as initial-value problems. Due to the singular nature of the stagnation point, where $q = \Oh(w^{1/2})$, we use the coordinate transformation $s(w) = w^{1/2}$, and solve the associated differential equations in $s$. The asymptotic behaviour \eqref{q0shipspec} is used to provide the initial value for $q$ at a point near $s = 0$. In the simulations, we typically used $s = 10^{-5}$, and the resultant amplitudes are verified to be independent of the initial condition.

The combined results are shown in Fig.~\ref{fig:numerics}. The leading-order asymptotic approximations fit the data closely, and we see that both the model with simplified nonlinear ($N = 2^*$, shown as circle) and linear ($N =2$, shown as stars) formulations duplicate the requisite behaviours reviewed in Table~\ref{tab:models}.

\section{Discussion} \label{sec:discuss}

Throughout our analysis of \S\ref{sec:reduce} to \S\ref{sec:numerics}, we have chosen to stray from Tulin's formulation, which is encapsulated in the study of \eqref{bern3}. This mode of presentation was out of necessity; while the broad outline of Tulin's reduction is ultimately correct, the use of the unknown $\Q$-function renders the equation impractical for most applications. Moreover, the \cite{davies_1951_the_theory} substitution \eqref{sinreduce} and subsequent truncation of the nonlinear $\P$ does not make it clear what inaccuracies are introduced by the reduction process (we have concluded, for example, that the pre-factor of the wave will be incorrect). We have provided an extended discussion of Tulin's $\Q$-function in Appendix~\ref{sec:Qfunc}, and the connections with our own formulation in \S\ref{sec:tulinconnect} and Appendix~\ref{sec:tulinconnect2}.

While Tulin's work may have been unappreciated since its inception, the work was, in fact, ahead of its time. Indeed, the proposal of the complex-variable reduction of the water-wave equations, and the subsequent simplification of the Hilbert transform would anticipate many of the more sophisticated asymptotic approaches that would later develop independently (\emph{e.g.} in \citealt{chapman_2006_exponential_asymptotics}). As we have reviewed in \S\ref{sec:otherworks}, others have proposed integral formulations of the low-Froude problem (see \emph{e.g.} the collection of models in \citealt{doctors_1980_comparison_of}), but such models typically depended on \emph{ad-hoc} linearizations of the two-dimensional potential flow equations. 

Also at the forefront of our motivation was to better understand Tuck's series of papers \citep{tuck_1990_water_non-waves, tuck_1991_ship-hydrodynamic_free-surface, tuck_1991_waveless_solutions}, positing simplified toy models that could eliminate the Hilbert transform, while preserving essential details of the waves. During the brief exchange between Tulin and Tuck, as quoted in \S{1.2}, Tuck had indicated that it was unclear whether his reductions were related, via the complex-plane, with Tulin's model. The answer is that they are indeed related. 

Both Tulin and Tuck's formulations are intended to be (but are ultimately incomplete) truncated models valid at low Froude numbers. Both studies attempt to produce an approximate wave solution independent of the Hilbert transform. Through our corrected reduction \eqref{simpsys} and study using steepest descents, we were able to explain why the Hilbert transform was crucial in some cases (determining $I_{\text{endpoints}}$), but negligible in others (determining $I_{\text{exp}}$). This idea of being able to study and visualize wave-structure interactions using the method of steepest descents is a powerful one, and we believe that it has wide applicability to further developing analytical theory for more complicated geometries than the ones we have considered here.

\subsection{Which model is correct?}

Another main result of our work relates to the presentation of Table~\ref{tab:models}, which unifies the various truncated models under consideration. We have shown that in the limit $\ep \to 0$, the exponentially small water waves are of the form $\mathcal{A}F(\phi)\e^{-\chi/\ep}$ plus its complex conjugate. 

In order to obtain the correct $\chi$ (and thus preserve some of the most important aspects), we can solve the one-term truncated model
\begin{equation}
  \ep \bq' + \chi'(w)\bq \sim -\ep q_0.
\end{equation}
This is the simplest reduction of the water wave problem. Despite the fact that it will incorrectly predict $\mathcal{A}F(\phi)$, it still serves as a useful toy model since the steepest descent argument remains unchanged.

If we wish to obtain the correct functional form of $F(\phi)$, then we can instead solve the equation,
\begin{equation}
\ep \bq' + \biggl[ \chi'(w) + \ep P_1'(w) + \Oh(\ep^2) \biggr]\bq 
= R(w; \bH[\bt]),
\end{equation}
where $R$ will be truncated to $\Oh(\ep^2)$ error, as detailed in Table~\ref{tab:models}.

It is difficult to propose any reduced equation that allows us to determine the correct numerical pre-factor, $\mathcal{A}$, since this involves inclusion of terms up to optimal truncation in $q_r$ and $\theta_r$. Since this optimal truncation term tends to infinity as $\ep \to 0$, then we must do as we have done in \emph{e.g.} \eqref{nearR}, and approximate $R$ using its divergent form. This is the connection with previous approaches that have used exponential asymptotics. The \emph{correct} model---that is to say, the one that predicts the leading-order exponential, up to the numerical pre-factor---does indeed require accounting for the divergent nature of the asymptotic approximations. 

\subsection{Reviewing Tulin's low speed comments} \label{sec:tulinlowspeed}

Tulin's paradoxical comment regarding the validity of the low speed limit, quoted on p.~\pageref{tulinquote} of our introduction, will be resolved once the specific problem geometry and integrand functions of \eqref{finalint} is considered using the method of steepest descents. 

In this case, we have seen through the methodology of \S\ref{sec:steep} that, for the case of ship flow, in the limit $\ep \to 0$, the wave contributions arise from deformation of the integration contours about the saddle points, which corresponds to the corner of the ship geometry. While the generated waves are unbounded near these critical points, they remain exponentially small order everywhere along the free surface. Indeed, it was precisely this argument that had allowed us to neglect the Hilbert transform. Therefore, there is no issue with taking the $\ep \to 0$ limit within the physical domain.

Tulin's comment, however, has an important consequence in the case of bow flows. In this case, we see that as the solution is analytically continued in the direction of the bow, the generated exponential, of order $\e^{-\chi/\ep}$, will tend to infinite amplitude at the stagnation point. Thus there is no bounded solution in the $\ep \to 0$ limit for bow flow. This situation was formally discussed in \cite{trinh_2011_do_waveless}, but it has been known [see \emph{e.g.} \cite{vanden-broeck_1977_computation_of}] that the numerical problem is not well-posed when a stagnation point attachment is assumed for the case of incoming flow. We have thus demonstrated this, here, for the limiting case of $\ep \to 0$. 

\subsection{Applicability of reduced models to further studies}

In an age where there are a bevy of tools and packages that can perform full numerical computations of the nonlinear water-wave problem, the reader may be justified in wondering whether there is still applicability in studying the significance of historical works by \cite{tulin_1982_an_exact}, \cite{tuck_1990_water_non-waves, tuck_1991_ship-hydrodynamic_free-surface, tuck_1991_waveless_solutions}, and this paper itself. As summarized by Table~\ref{tab:models}, the differences between various truncated models are subtle, and given the current state of computation, it seems more difficult to unravel such subtleties than it would be to to solve the full model. 

However, while methods of computation have improved significantly since Tulin's 1982 paper, many theoretical aspects of free-surface wave-body flows are still a mystery, as evident by the review in \cite{tulin_2005_reminiscences_and}. For example, there is virtually no analytical theory that can distinguish between waves produced by surface-piercing bodies with sudden angular changes (corners) versus bodies that are smooth [\emph{c.f.} the discussion in \cite{trinh_2014_the_wake}]. The classic treatments using linearized theory, as it appears in \emph{e.g.} \cite{kostyukov_1968} and \cite{wehausen_1973}, are limited to asymptotically small bodies rather than the bluff bodies we consider in this paper. In a forthcoming work, we will demonstrate how the steepest descent methodology developed in this paper, can be applied to the study of smooth-bodied obstructions. 

The techniques and reductions presented in this paper, along with further developments in the theory of exponential asymptotics, provides hope that analytical progress can be made on the subject of time-dependent and three-dimensional wave-body problems [\emph{c.f.} recent work by \cite{howls_2004}, \cite{chapman_2005}, \cite{lustri_2014a}, \cite{bennett_2015_exponential_asymptotics} on this topic].

\begin{appendix}
\section{Form of the divergence} \label{sec:divergence}

The individual components, $Q$, $\chi$, and $\gamma$, that make up the factorial-over-power divergence in \eqref{qt_fact} can be derived by examining the governing equations at $\Oh(\ep^n$). In the limit $n \to \infty$, the leading-order contribution gives the \emph{singulant}, $\chi$, 
\begin{equation}
  \dd{\chi}{w} = \frac{\im \j \k}{q_0^3},
\end{equation}
and since $\chi = 0$ at the singularities, we write
\begin{equation}
  \chi = \im \j \k \int_{w_0}^w \frac{\de{\varphi}}{q_0^3(\varphi)},
\end{equation}
where $w = w_0$ is a particular singularity of the leading-order solution. 

Similarly, it can be shown that
\begin{equation} \label{Q}
Q(w) = \frac{\Lambda}{q_0^2(w)} \exp \left[ 3 \im \j \k \int_{w^*}^w \frac{q_1(\varphi)}{q_0^4(\varphi)} \, \de{\varphi}\right],  
\end{equation}
where $\Lambda$ is a constant of integration and $w^*$ is an arbitary point chosen wherever the integral is defined. The pre-factor $\Theta$, is then related to $Q$ using
\begin{equation}
  \Theta = (\i \j\k) q_0 Q.
\end{equation}
The value of $\gamma$ is derived by matching the local behaviour of $q_n$ in \eqref{qt_fact} with leading-order $q_0$ near the singularity, $w = w_0$. If we assume that $q_0 \sim c (w - w_0)^\alpha$ near the singularity, then 
\begin{equation}
  \gamma = -\frac{6\alpha}{1 - 3\alpha}.
\end{equation}

For most nonlinear problems, the value of $\Lambda$ in \eqref{Q} embeds the nonlinearity of the governing equations near the singularity, and must be found through a numerical solution of a recurrence relation. A detailed derivation of the above quantities, including numerical values of $\Lambda$, can be found in \S{3.1} of \cite{chapman_2006_exponential_asymptotics} and \S{4} of \cite{trinh_2011_do_waveless}. We also refer the reader to more general reviews of exponential asymptotics by \cite{olde-daalhuis_1995_stokes_phenomenon}, \cite{boyd_wnlsw}, and \cite{costin_book}.

\section{The limitations of the $\Q$-function} \label{sec:Qfunc}

The most difficult aspect of Tulin's work concerns Section VI of the manuscript, which seeks to understand the nature of the analytically continued function, $\Q(w)$. We will attempt to follow the same argument as Tulin (with adjustments for changes in notation and flow geometry). 

\begin{figure} \centering 
\vspace*{1.0\baselineskip}
\includegraphics[width=0.8\textwidth]{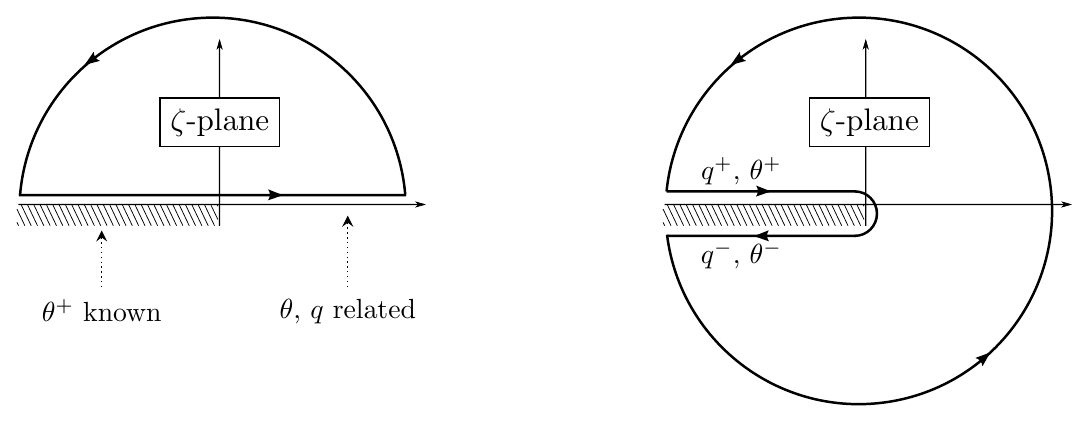}
\caption{(Left) The Cauchy contour used to derive the boundary integral formulation \eqref{firstbdint}; (right) The Cauchy contour used in Tulin's $\Q$-function. \label{fig:cauchy}}
\end{figure}

Tulin had split the form of $\Q$ into a contribution from the uniform flow and a contribution from the geometry. In his notation, our $\Q(w)$ is equal to $\i\j - Q(w)$. The situation of an imposed pressure distribution was also considered in his work, but we shall ignore this effect. Tulin had then written $\Q$ in terms of a Cauchy integral over the solid boundary and its image reflected about the free surface in the potential plane. This is analogous to applying Cauchy's integral theorem to either $G$ or the hodograph variable \eqref{hodograph} along a counterclockwise circular contour of radius $R \to \infty$ with a slit about the negative real axis (see Fig.~\ref{fig:cauchy}, right). This yields
\begin{equation} \label{tulinCauchy}
    \log q(\zeta) - \i \theta(\zeta) = \frac{1}{2\pi \i} \left(\int_\text{circle} + \int_\text{slit}\right) \left[\frac{\log q - \i \theta}{t - \zeta}\right] \, \de{t},
\end{equation}
where the integral along the outer circle tends to zero as $R \to \infty$ by the boundary conditions and the integral over the slit involves $q^{\pm}$ and $\theta^\pm$, the limiting values from the upper or lower half-planes. However, in \eqref{tulinCauchy}, only the values of $\theta$ on one side are known (being the physical angle of the solid boundary). For instance, in the case of a step, $\theta^+(t)$ is given by \eqref{steptheta}. The problem, however, is that the other values $\theta^-(t)$, $q^+(t)$, and $q^-(t)$ are only known through analytic continuation, and it is impossible to go further with \eqref{tulinCauchy} without additional information.

For most practical implementations, it is preferable to instead apply Cauchy's Theorem to an integral over the solid boundary and the free surface (see Fig.~\ref{fig:cauchy}, left), which was the formulation in \S\ref{sec:bdint}. Bernoulli's equation is required in order to provide a relationship between $q$ and $\theta$ on the free surface, and thus close the system. 

Tulin had posited that the value of $\Q$ might be found through a theoretically posited surrogate body whose singularities on the physical boundary alone would generate the flow (rather than the physical boundary and its reflected image). This appears as eqn (60) in his work. However, for a given physical geometry (\emph{e.g.} for the step and ship geometries of \S\ref{sec:form}), it is unclear how this surrogate body could ever be determined in an \emph{a priori} fashion.   

Thus in the end, Tulin's $\Q(w)$ should rather be written as $\Q(w; q, \theta)$, as it involves the solution itself. This creates a problematic argument if the intention is to treat \eqref{bern3} as an ordinary differential equation to be integrated exactly, for the solution appears on both sides of the formulation. Indeed, this is precisely the issue that Tuck (\S\ref{sec:introtuck}) had wrestled with, in seeking a reduction of the global Hilbert transform operator. 

In the approach to follow, we will resume our study of the formulation in \S\ref{sec:analytic}, and we will return to discuss the connection to Tulin's work in \S\ref{sec:discuss}.

\section{Connection to Tulin's formulation} \label{sec:tulinconnect2}

\noindent As explained in \S\ref{sec:tulinconnect}, we have chosen to stray from Tulin's formulation, which uses the combined analytic function, $G = (q\e^{-\im\j\theta})^3$, and the unknown right hand-side, $\Q$. Our formulation separates the analytic continuations of the $q$ and $\theta$ variables, and is self contained. In contrast, Tulin's formulation requires the specification of the $\Q$-function, which requires additional information. We now wish to show how Tulin's equation \eqref{tulinagain} is related to the equation for the exponential in \S\ref{sec:hankel}, given by 
\begin{equation} \label{expeqagain}
  \ep \bq' + \Bigl[ \chi' + \ep P_1' \Bigr]\bq \sim -\ep^N q_{N-1}', 
\end{equation}
that is, the reduced integro-differential model \eqref{simpsyseq} with the right hand-side \eqref{nearR}. 

In order to relate the two formulations, we multiply \eqref{tulinagain} through by $G$ and obtain
\begin{equation} \label{tulinmult}
  \ep \dd{G}{w} - \im \j + (\P G) = -\Q G.
\end{equation}
We expand the unknown functions in the above equation into a regular perturbation expansion and an error term, using
\begin{equation}
  G = G_r + \bG, \qquad (\P G) = (\P G)_r + \bPG, \qquad \Q = \Q_r + \bQ,
\end{equation}
where for simplicity, we expand the product $(\P G)$ rather than the individual factors. Substitution into \eqref{tulinmult} gives
\begin{equation} \label{tulinmult2}
\ep \dd{\bG}{w} + \Q_r \bG = -\Ebb + \Oh(\bar{\Q}, \bPG),
\end{equation}
where we have introduced the error in the Bernoulli equation, 
\begin{equation} \label{Ebb}
  \Ebb = \ep \dd{G_r}{w} - \im \j + (\P G)_r + \Q_r G_r,
\end{equation}
which can be compared to \eqref{Ebern}. The expansion of $G = (G_0 + \ep G_1 + \ldots) + \bG$, can also be written in terms of the expansions for $q$ and $\theta$. Using $G = (q\e^{-\im\j\theta})^3$, and expanding, we find
\begin{subequations} \label{G0G1Gb}
\begin{gather}
G_0 = q_0^3, \qquad
G_1 = 3 q_0^2 q_1 - 3 \im \j q_0^3 \theta_1, \label{G0G1} \\
\bG = \Bigl[3 q_0^2 \bq - 3 \im \j q_0^3 \bt\Bigr] + \ep \Bigl[ -9 \im \j q_0^2 q_1 \bt 
- 9 \im \j q_0^2 \theta_1 \bq + 6 q_0 q_1 \bq - 9 q_0^3 \theta_1 \bt \Bigr] + \ldots. \label{Gbexpand}
\end{gather}  
\end{subequations}

Following the discussion in \S\ref{sec:tulinconnect} and Appendix~\ref{sec:Qfunc}, we emphasize that $q_0$, $q_1$, $\theta_1$, and the low-order terms are derived independently from \eqref{tulinmult}---that is, the single equation for $G$ is insufficient to close the system without inclusion of the Hilbert transform. Instead, we assume that the low-order terms in \eqref{G0G1Gb} are known and that \eqref{tulinmult} provides an equation for $\Q$. The regular part of $\Q$, given by $\Q_r$, follows from expansion of the left hand-side of \eqref{tulinagain}. Only including up to $\Oh(\ep^2)$ terms, we obtain 
\begin{equation} \label{Qr}
  -\Q_r = \ep \frac{G_0'}{G_0} - \frac{\im \j}{G_0} + \ep \frac{\im \j G_1}{G_0^2} + \Oh(\ep^2) = -\frac{\im\j}{q_0^3} - \ep \left[-\frac{3 q_0'}{q_0} - \frac{3 \theta_1}{q_0^3} - \frac{3 \im \j q_1}{q_0^4}\right] + \Oh(\ep^2). 
\end{equation}

We also note that in the limit $\ep \to 0$, the optimal truncation point of the regular series expansion tends to infinity, $N \to \infty$, and the error in Bernoulli's equation is replaced by the divergent term
\begin{equation} \label{Ebb2}
  \Ebb \sim \ep^N G_{N-1}' \sim \ep^N q_{N-1}', 
\end{equation}
which is analogous to the argument leading to \eqref{nearR}. The result now follows by using \eqref{asym1} for $q_1$ and $\theta_1$, \eqref{Gbexpand} for $\bG$, \eqref{Qr} for $\Q_r$, and \eqref{Ebb2} for the right hand-side of \eqref{tulinmult2}. We are left with
\begin{equation}
  \ep \bq' + \left[ \frac{\im\j}{q_0^3} + \ep \left( \frac{2 q_0'}{q_0} - \frac{3\im\j q_1}{q_0^4}\right)\right] \bq \sim -\ep^N q_{N-1}',
\end{equation}
or \eqref{expeqagain}, as desired. Thus we have shown how Tulin's formulation will exactly preserve the exponentially small surface waves. Derivation of the full relationship of Tulin's equation to the full system \eqref{simpsys} can be similarly done, but the algebra (in expanding $\Q$ and $G$, and returning to the formulation with the embedded Hilbert transform) becomes unwieldy.

\end{appendix}
 

\providecommand{\noopsort}[1]{}

\end{document}